\documentclass[twocolumn]{aastex701}

\usepackage{verbatim}

\submitjournal{ApJ}

\shorttitle{ZSAD in CEERS and NGDEEP}
\shortauthors{Surakan et al.}

\begin{document}

\title{Evaluating Zero-Shot Anomaly Detection for Identifying Unusual Galaxy Morphologies in CEERS and NGDEEP}

\author[0009-0001-8122-2785, gname=Tai, sname=Surakan]{Tai B. Surakan}
\affiliation{Laboratory for Multiwavelength Astrophysics, School of Physics and Astronomy, Rochester Institute of Technology, 84 Lomb Memorial
Drive, Rochester, NY 14623, USA}
\email[show]{taibsurakan@gmail.com}  

\author[orcid=0000-0001-9187-3605,gname=Jeyhan, sname=Kartaltepe]{Jeyhan S. Kartaltepe}
\affiliation{Laboratory for Multiwavelength Astrophysics, School of Physics and Astronomy, Rochester Institute of Technology, 84 Lomb Memorial
Drive, Rochester, NY 14623, USA}
\email{jeyhan@astro.rit.edu}

\author[0000-0003-3835-9898]{Rohan Pattnaik}
\affil{Johns Hopkins University, Department of Physics \& Astronomy, 3400 N Charles Street, Baltimore, MD 21218, USA}
\email{rpattna1@jh.edu}

\author[0009-0009-4635-9442]{Clive K. Binu}
\affiliation{Laboratory for Multiwavelength Astrophysics, School of Physics and Astronomy, Rochester Institute of Technology, 84 Lomb Memorial
Drive, Rochester, NY 14623, USA}
\email{}

\author[0000-0002-1803-794X]{Isa G. Cox}
\affiliation{Laboratory for Multiwavelength Astrophysics, School of Physics and Astronomy, Rochester Institute of Technology, 84 Lomb Memorial
Drive, Rochester, NY 14623, USA}
\email{}

\author[0000-0003-4512-8705]{Tiger Y. Hsiao}
\affiliation{Cosmic Frontier Center, The University of Texas at Austin, Austin, TX 78712, USA}
\email{}

\author[0000-0003-1581-7825]{Ray A. Lucas}
\affiliation{Space Telescope Science Institute, 3700 San Martin Drive, Baltimore, MD 21218, USA}
\email{lucas@stsci.edu}

\author[0000-0001-9879-7780]{Fabio Pacucci}
\affiliation{Center for Astrophysics, 60 Garden Street, Cambridge, MA 02138, USA}
\email{fabio.pacucci@cfa.harvard.edu}

\author[0000-0001-5930-0532]{Gregory Troiani}
\affiliation{Department of Physics and Astronomy, University of Kansas, Lawrence, KS 66045, USA}
\email{gtroiani@ku.edu}

\author[0000-0003-3466-035X]{{L. Y. Aaron} {Yung}}
\affiliation{Institute of Astronomy and Department of Physics, National Tsing Hua University, Hsinchu 30013, Taiwan}
\email{yung@phys.nthu.edu.tw}

\begin{abstract}
Galaxy morphology out to $z\sim 6$ provides insight into galaxy formation and evolution, but identifying rare or structurally complex systems is difficult due to observational limits and the scale of modern imaging surveys. Here, we apply AdaCLIP, a Zero-Shot Anomaly Detection (ZSAD) model built upon the Contrastive Language-Image Pretraining (CLIP) model and adapted using non-astronomical anomaly-detection datasets, to identify structurally unusual galaxies without galaxy-specific anomaly-detection training or morphological labels. We extract anomaly scores and 768-dimensional feature embeddings from 38,853 galaxies in the Cosmic Evolution Early Release Science (CEERS) and Next Generation Deep Extragalactic Exploratory Public (NGDEEP) surveys, and compare the resulting anomaly scores with parametric \texttt{GALFIT} and nonparametric \texttt{statmorph} measurements. The feature embeddings reveal structure associated with galaxy morphology and observational properties, while anomaly scores show modest trends with structural measures, including asymmetry, concentration, $M_{20}$, and size, but weaker or less consistent trends with Sérsic index, axis ratio, smoothness, and Gini coefficient. These results highlight the potential of ZSAD as a scalable tool for identifying unusual galaxy morphologies in large astronomical imaging surveys, including those anticipated from the Nancy Grace Roman Space Telescope.
\end{abstract}


\section{Introduction}
\label{sec:introduction}
Galaxies exhibit a wide range of structural properties, including smooth, spheroidal morphologies, disk-dominated systems, and highly irregular structures. This diversity motivated the development of the Hubble Sequence \citep{Hubble1926}, which classifies galaxies into elliptical, spiral, and irregular systems based on their observed structure. Although originally conceived as a descriptive framework, galaxy morphology is now understood to reflect the underlying physical processes governing stellar populations, star formation, dynamical state, and interactions (e.g., \citealp{Roberts1994}; \citealp{Kennicutt1998}; \citealp{Conselice2014}). As a result, morphological measurements are a key observational probe of galaxy evolution, linking structure to physical processes such as gas accretion, mergers, and the growth of supermassive black holes (e.g., \citealp{Lotz2008}; \citealp{Dekel2009}; \citealp{Conselice2014}).

The launch of the Hubble Space Telescope (HST) \citep{Bahcall1989} in 1990 marked a turning point in astronomy. Its observations revealed a much higher prevalence of irregular and asymmetric galaxies at high redshift compared to the local universe \citep{Abraham1996}, including populations of clumpy and chain-like galaxies \citep{Elmegreen2004}. Subsequent studies also demonstrated the evolution of more organized galactic structures, such as the decreasing fraction of barred spiral galaxies with increasing redshift \citep{Sheth2008}. Landmark surveys, such as the Hubble Deep Field (HDF) \citep{Williams1996}, the Great Observatories Origins Deep Survey (GOODS) \citep{Giavalisco2004}, the Hubble Ultra Deep Field (HUDF) \citep{Beckwith2006}, and the Cosmic Assembly Near-Infrared Deep Extragalactic Legacy Survey (CANDELS) (\citealp{Grogin2011}; \citealp{Koekemoer2011}), revealed a rich diversity of galaxies across cosmic time (e.g., \citealp{vanderwel2014}; \citealp{Kartaltepe2015}), with the deepest observations reaching epochs as early as $\sim 500$ million years after the Big Bang \citep{Bouwens2011a}.

Despite its transformative contributions, HST observations at high redshift are limited by the wavelength coverage of its WFC3/IR broad-band imaging filters, which extend to only $1.6$ $\mu$m. Beyond $z \gtrsim 3$, rest-frame optical emission that traces the underlying stellar mass distribution is redshifted beyond this limit, restricting morphological studies to rest-frame ultraviolet (UV) light (\citealp{Bell2001}; \citealp{Bouwens2015}). Because rest-frame UV emission is dominated by short-lived star-forming regions \citep{Casey2014} and is strongly impaired by dust attenuation and stellar feedback processes (e.g., \citealp{Bell2002, Erb2015, Salim2020}), it provides an incomplete and often misleading or biased view of galaxy structure, complicating the identification of disks, bulges, and merger-driven features.

Building on HST's legacy, the James Webb Space Telescope (JWST) \citep{Gardner2006} has transformed the study of galaxy morphology at high redshift. Optimized for observations in the near- and mid-infrared, it can probe wavelengths less affected by dust attenuation and more representative of underlying stellar mass distributions. JWST provides a substantially clearer view of galaxy structure at high redshift, including during cosmic noon ($1.5 \leq z \leq 3$), when the universe experienced peak rates of star formation and black hole growth \citep{Delvecchio2014, Madau2014}. This epoch marks a pivotal stage in galactic evolution, during which disks assembled and stabilized, bulges grew, mergers reshaped galaxies, and morphological diversity increased substantially (e.g., \citealp{Dekel2009, ForsterSchreiber2009, Conselice2014, Kartaltepe2015}). Understanding galaxy structure before, during, and after cosmic noon is therefore essential for tracing how early galaxies evolved into the ordered Hubble-type systems observed in the local universe.

The enhanced capabilities of JWST have also introduced new challenges. The scale and complexity of modern imaging surveys, together with the growing and anticipated datasets from wide-field facilities such as the Nancy Grace Roman Space Telescope \citep{Spergel2015}, the Vera C. Rubin Observatory \citep{Ivezic2019}, and Euclid \citep{Mellier2025}, require increasingly automated and scalable analysis techniques. Citizen science efforts such as Galaxy Zoo \citep{Lintott2008} demonstrated that visual classifications can successfully characterize galaxy morphology on a large scale \citep{Willett2013}, providing morphological classifications for over one million galaxies \citep{Masters2020}. Subsequent work has also sought to reduce systematic biases in human classifications through statistically calibrated frameworks while retaining expert or volunteer classifications as the foundation of the analysis \citep{Lambrides2021}. However, visual approaches remain time-intensive, subject to classifier inconsistencies, and difficult to scale to the data volumes expected from next-generation surveys. Additionally, quantitative morphological diagnostics, including parametric Sérsic-profile fitting \citep{Sérsic1963} and nonparametric statistics (e.g., \citealp{Conselice2003, Lotz2004}), have been extensively developed to characterize galaxy structure. However, these measurements can remain sensitive to observational effects, such as image depth and spatial resolution \citep{Sazonova2024}, complicating efforts to characterize the structural complexity and diversity of galaxies at intermediate to high redshift (e.g., \citealp{Kartaltepe2010, Kartaltepe2023}). These limitations have motivated the development of machine learning (ML) techniques capable of identifying morphological patterns, rare systems, and anomalous structures directly from large imaging datasets.

For astronomical images, ML techniques can process large volumes of data much more efficiently than traditional visual classification. Supervised techniques are typically trained on carefully labeled datasets that identify pretrained morphological classes, such as spiral or elliptical galaxies (e.g., \citealp{Dieleman2015, HuertasCompany2015, DominguezSanchez2018, Walmsley2022}). However, applying these approaches to galaxies at higher redshift is challenging, where morphologies are more diverse and often differ substantially from those observed in the local universe. While recent efforts have begun to provide visual morphological classifications for thousands of JWST galaxies (e.g., \citealp{Troiani2026}), large, reliably labeled training sets spanning this broader range of galaxy structures are not readily available, and individual conventional morphological classes may not adequately capture their structural complexity \citep{Kartaltepe2023}. These limitations motivate the use of alternative methods that can characterize galaxy structure without relying on predefined morphological labels.

Unsupervised methods aim to identify intrinsic structure in data to group or embed objects based on similarity without predefined class labels \citep{Feigelson2012, Baron2019, Kolesnikov2024}. Such approaches have been successfully applied to galaxy morphology, identifying structurally distinct populations in both the Sloan Digital Sky Survey and JWST observations \citep{Cheng2021, Tohill2024}. Zero-shot methods take a different approach, leveraging pretrained representations to generalize to unseen classes or tasks without requiring additional training data \citep{Lampert2009, Frome2013, Radford2021}. In the context of galaxy morphology, these approaches offer opportunities to explore structural diversity and identify rare systems, structural outliers, and previously uncharacterized morphological features in large imaging surveys.

One application of zero-shot learning is zero-shot anomaly detection (ZSAD), which aims to identify unusual visual patterns in image categories not encountered during anomaly-detection training. Recent ZSAD approaches leverage pretrained vision-language models (VLMs) which encode visual and textual information into comparable representations, to distinguish normal and anomalous visual appearances \citep{Jeong2023, Zhou2023}. In particular, AdaCLIP \citep{Cao2024} adapts these representations using labeled examples of normal and anomalous appearances from auxiliary datasets, such as industrial and medical imagery, without requiring training examples from the target category. AdaCLIP produces an anomaly score between 0 and 1 based on an image's relative similarity to textual descriptions of normal and abnormal states, with higher scores indicating greater similarity to the latter. This provides an opportunity to investigate whether anomaly-detection representations adapted using non-astronomical datasets can identify unusual galaxy structures without requiring galaxy-specific training or morphological labels.

In this work, we developed an AdaCLIP-based anomaly detection pipeline for galaxy imaging spanning $0 < z < 6$. We are particularly interested in identifying candidate morphological anomalies, including galaxies exhibiting disturbed, asymmetric, irregular, or multi-component structures that may be associated with physical processes such as mergers and interactions. We generate anomaly scores and embedding spaces to investigate morphological similarities and identify galaxies exhibiting these unusual structural features in JWST observations. We extend this framework to a quantitative morphological analysis of galaxies in the Cosmic Evolution Early Release Science (CEERS) \citep{Finkelstein2025} and Next Generation Deep Extragalactic Exploratory Public (NGDEEP) \citep{Bagley2024} surveys. We incorporate parametric morphology measurements derived from \texttt{GALFIT} modeling \citep{McGrath2026} for the CEERS sample and also compute nonparametric structural parameters using \texttt{statmorph} \citep{RodriguezGomez2019} to characterize galaxy structure using established morphological diagnostics. We then compare these morphology measurements to the AdaCLIP anomaly scores, using the embedding space as a visual tool to assess morphological similarity. This comparison provides a direct evaluation of how effectively the AdaCLIP anomaly score captures established morphological properties, particularly for faint and high-redshift galaxies where traditional morphological measurements may become less reliable.

This paper is organized as follows. Section~\ref{sec:data} describes data and imaging reduction procedures. Section~\ref{sec:anomaly_detection_pipeline} introduces the AdaCLIP-based anomaly detection pipeline and its outputs. Section~\ref{sec:galaxy_morphology} reviews traditional parametric and nonparametric morphology metrics and describes additional sample selection criteria. The results are presented in Section~\ref{sec:results} and their implications are discussed in Section~\ref{sec:discussion}. Finally, we summarize our findings in Section~\ref{sec:summary_conclusion}. All magnitudes are reported in the AB system \citep{Oke1983}. We also assume the following cosmological parameters: $H_{0} = 70$ km s$^{-1}$ Mpc$^{-1}$, $\Omega_{tot}=1$, $\Omega_{\Lambda}=0.7$, and $\Omega_{M}=0.3$.

\section{Data}
\label{sec:data}
We use JWST/NIRCam \citep{Horner2004} imaging data from the CEERS and NGDEEP surveys to extract galaxy cutouts for morphological analysis. The CEERS survey provides deep near-infrared imaging of distant galaxies in the well-studied Extended Groth Strip (EGS) field \citep{Davis2007}. The survey is designed to investigate galaxy formation and evolution across cosmic time, from the epoch of reionization through cosmic noon, with particular emphasis on how galaxies assemble their stellar populations, grow disks and bulges, and undergo morphological transformation through mergers and accretion processes \citep{Finkelstein2025}. These observations also enable rest-frame UV-to-optical studies of galaxy structure, providing insight into how morphology evolves with redshift and physical conditions such as star formation, metallicity, and dust attenuation. 

For this study, we use the calibrated JWST/NIRCam mosaics in all six broadband filters common to CEERS and NGDEEP (F115W, F150W, F200W, F277W, F356W, and F444W; \citealp{Bagley2023, Bagley2024, Cox2026}), providing multiwavelength coverage for investigating galaxy structure across cosmic time. Among these filters, F277W reaches the greatest median $5\sigma$ point-source depth in CEERS, approximately 29.3 mag \citep{Cox2026}. For CEERS, we use the v1.2 photometric catalog to define the galaxy sample and the v1.0 NIRCam mosaics to extract image cutouts. Specifically, we use the catalog source identifiers and celestial coordinates to locate each galaxy, Kron aperture measurements to determine adaptive cutout dimensions, and catalog-derived quality flags, photometric redshifts, flux measurements, and flux uncertainties to apply the selection criteria described in Section~\ref{subsec:source_selection}. The pixel values analyzed in this work are extracted directly from the calibrated JWST science images.

The NGDEEP survey leverages JWST's parallel observing strategy to obtain ultra-deep NIRCam imaging (NGDEEP-NRC) of the Hubble Ultra Deep Field parallel field (HUDF-Par2; \citealp{Oesch2009, Bouwens2011b}), a well-studied region with extensive ancillary observations \citep{Bagley2024}. Designed to probe the faintest and highest-redshift galaxy populations accessible to JWST, NGDEEP achieves substantially greater sensitivity than CEERS through longer integration times over a smaller survey area. In the F277W filter, \citet{Bagley2024} report an expected $5\sigma$ resolved-source limiting magnitude of approximately 30.9 for the full NGDEEP-NRC survey. For this work, we utilize the calibrated NGDEEP-NRC NIRCam mosaics and the corresponding NGDEEP source catalog (Finkelstein et al., in preparation), which provides photometric measurements and source properties for objects identified in the survey.

\subsection{Source Selection}
\label{subsec:source_selection}
The CEERS NIRCam catalog \citep{Cox2026} contains 87,938 objects, while NGDEEP-NRC \citep{Bagley2024} contains 17,177, yielding a combined sample of 105,115 detected sources. We perform an initial quality-control step to remove unreliable detections. Specifically, we exclude sources identified as stars, objects located in catalog-defined bad image regions, and sources with invalid (e.g., NaN values) photometric measurements, including fluxes in any of the six NIRCam filters or photometric redshifts. These criteria remove objects with incomplete or unreliable measurements that could bias subsequent morphological analysis, reducing the combined sample to 90,443 sources.

We next apply a signal-to-noise ratio (SNR) threshold of SNR $ > 10$ using the F277W Kron flux and its empirical uncertainty. This threshold retains well-detected sources with sufficient signal for reliable morphological measurements and removes the faintest detections, whose structure is poorly resolved and more likely to be dominated by noise or residual artifacts. We use the F277W filter because it provides the deepest NIRCam observations in the CEERS dataset \citep{Finkelstein2025, Cox2026} while also serving as a common reference band across both CEERS and NGDEEP. Although F115W reaches greater depth in NGDEEP \citep{Bagley2024}, adopting F277W ensures a consistent selection across the combined catalogs and provides long-wavelength measurements that are better suited for the high-$z$ galaxies in our sample. This selection removes any low-confidence detections and spurious sources in the combined CEERS+NGDEEP dataset, reducing the total number of objects to 44,700.

Next, we limit sources to an F277W magnitude range of $20 < m_{\rm{F277W}} < 28$, leaving a combined total of 40,095 objects. While the SNR threshold ensures reliable photometric detections, the additional magnitude restriction establishes a consistent apparent-brightness range across CEERS and NGDEEP, which have different imaging depths. Together, these criteria are intended to produce a well-detected galaxy sample that retains a broad range of apparent brightnesses and morphological structures. Many canonical morphology diagnostics were originally developed and calibrated using relatively bright, well-resolved galaxy samples (e.g., \citealp{Conselice2003, Lotz2004}), and their behavior for faint, low-surface-brightness, or structurally irregular JWST systems may differ from that established regime (e.g., \citealp{Andrae2011, Conselice2014}). Including this broader magnitude range allows us to examine how traditional morphology measurements vary across the same population analyzed by AdaCLIP and, in turn, assess whether these measurements exhibit systematic relationships with the model's anomaly scores. This provides a basis for interpreting the ML-derived anomaly scores in terms of established structural properties while testing whether the model identifies morphological characteristics associated with increasingly unusual galaxy structure.

\begin{deluxetable*}{cccc}
\tablecaption{Data selection cuts}
\tablehead{
\colhead{Selection Step} & \colhead{CEERS} & \colhead{NGDEEP} & \colhead{Total}
}
\startdata
Initial sample & 87,938 & 17,177 & 105,115\\
Quality Control & 73,266 & 17,177 & 90,443\\
$\text{SNR}_{\text{F277W}}$ $> 10$ & 37,139 & 7,561 & 44,700\\
$20 < m_{\rm{F277W}} < 28$ & 35,527 & 4,568 & 40,095\\
$0 < z < 6$ & 34,445 & 4,503 & 38,853\\
\hline
\enddata
\tablecomments{Number of sources remaining after each selection step. NGDEEP does not have the quality control flags found in the CEERS photometric catalog. The final combined sample contains 38,853 sources.}
\label{tab:data_reduction}
\end{deluxetable*}

\begin{figure}[htb!]
\includegraphics[width=\columnwidth]{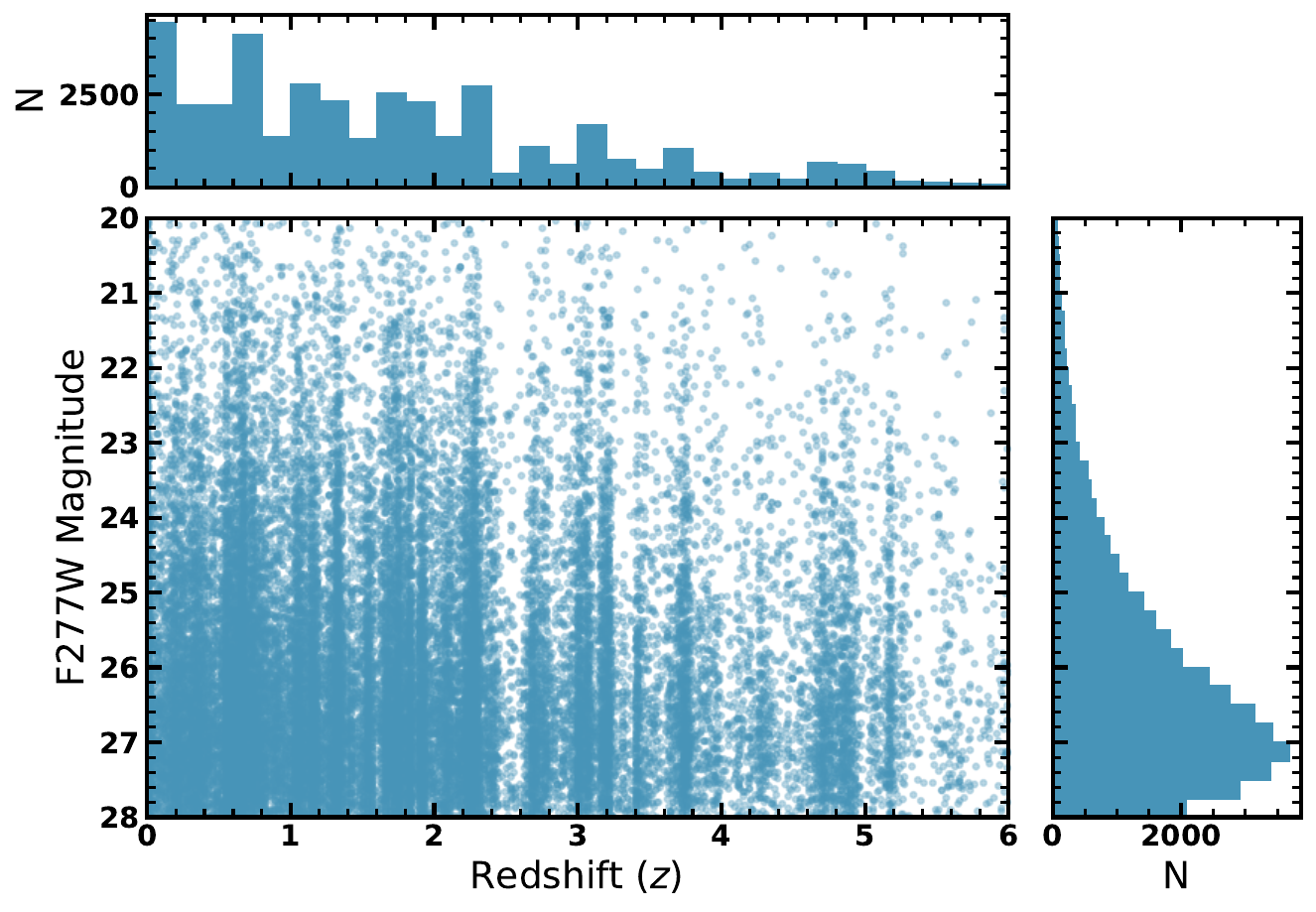}
\caption{F277W magnitude vs. redshift for the sample used in this paper. Above and to the right are the distributions of redshift and $m_{\rm{F277W}}$, respectively.}
\label{fig:sample_distribution}
\end{figure}

We also apply a photometric redshift selection criterion of $0 < z < 6$ using the catalog redshift estimates. This broad interval allows us to examine galaxies across multiple stages of cosmic evolution while focusing on populations whose morphological structures can be more readily characterized with the available NIRCam imaging. After applying all selection criteria, the final working sample contains 38,853 galaxies. The complete filtering sequence is summarized in Table~\ref{tab:data_reduction}, and Figure~\ref{fig:sample_distribution} shows the redshift and $m_{\rm{F277W}}$ distributions for the final combined sample.

\subsection{Image Creation}
\label{subsec:image_creation}
The CEERS and NGDEEP NIRCam mosaics have a common pixel scale of $0\farcs03$ $\mathrm{pixel}^{-1}$. We determine individual cutout sizes using the catalog parameters KRON\_RADIUS and A\_IMAGE, which define adaptive Kron apertures \citep{Kron1980}. We compute the effective semi-major axis using
\begin{equation}
    \text{semi-major axis} = \text{KRON\_RADIUS} \times \text{A\_IMAGE}\text{.}
\end{equation}
We multiply the resulting axis by a factor of 2.3 to define the cutout width and height, corresponding to a 15\% margin beyond the full Kron aperture for adaptively sized cutouts. We additionally enforce a minimum size of $64 \times 64$ pixels to retain surrounding structural information and potential nearby companions. This minimum is applied to approximately 98\% of the sample, resulting in a nearly uniform angular field of view of $1\farcs92$ per side for most galaxies.

\begin{figure}[htb!]
\includegraphics[width=\columnwidth]{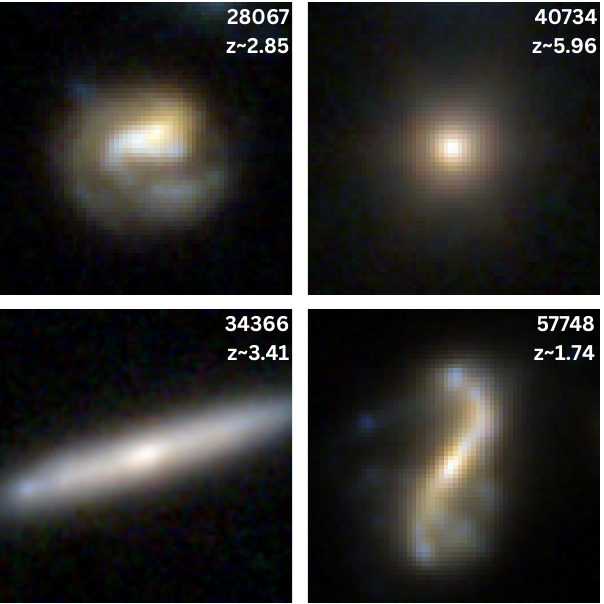}
\caption{Example CEERS galaxy color images constructed from the six NIRCam filters. Two filters are assigned to each color channel: F115W \& F150W are mapped to blue, F200W \& F277W to green, and F356W \& F444W to red. Cutouts are not shown at a common angular scale.}
\label{fig:galaxy_cutouts}
\end{figure}

After determining the cutout sizes, we combine the six filter cutouts into RGB images using \texttt{Trilogy}\footnote{https://github.com/dancoe/Trilogy} \citep{Coe2012}. The software package creates color images from multiband astronomical observations using nonlinear intensity scaling to enhance the visibility of faint structural features. We assign two filters to each color channel to capture complementary wavelength information: F115W \& F150W to blue, F200W \& F277W to green, and F356W \& F444W to red. We use identical configuration parameters for all images, including an upper-intensity saturation parameter of \texttt{satpercent = 0.0009}, corresponding to a nominal saturation of 0.0009\% of sampled pixels, and a color saturation factor of \texttt{colorsatfac = 1}. Given the relatively small dimensions of the cutouts, this saturation setting results in negligible upper-intensity clipping, minimizing information loss in the brightest pixels. Because we determine the cutout dimensions before image processing, we set \texttt{samplesize = 20000} pixels. Since all input cutouts contain fewer pixels than this limit, \texttt{Trilogy} uses the full extent of each cutout to determine its intensity scaling. For each cutout, \texttt{Trilogy} automatically determines the noise and saturation levels used for intensity scaling. The resulting RGB images serve as inputs to our anomaly-detection pipeline, as shown in Figure~\ref{fig:galaxy_cutouts}.

\section{Anomaly Detection Pipeline}
\label{sec:anomaly_detection_pipeline}
In this section, we present the anomaly detection pipeline used to identify morphologically unusual galaxy cutouts in the combined dataset. Section~\ref{subsec:ZSAD_motivation} discusses the motivation for ZSAD and its application to JWST imaging. Section~\ref{subsec:framework} introduces the AdaCLIP framework, including the model architecture, outputs, and initial configurations considered in this work. Finally, Section~\ref{subsec:model_selection} introduces the model selection criteria we use to determine which model configuration yields stable, physically meaningful anomaly rankings, and the reproducible pipeline is discussed in Section~\ref{subsec:pipeline}.

\subsection{Motivation for Zero-Shot Anomaly Detection}
\label{subsec:ZSAD_motivation}
Although an effective descriptor for nearby, well-resolved galaxies, the Hubble Sequence becomes increasingly inadequate for the irregular and disturbed morphologies revealed by deep JWST surveys (e.g., \citealp{Ferreira2022, Kartaltepe2023}). The scale and diversity of observed morphologies make comprehensive manual classification increasingly impractical. Additionally, modern supervised ML approaches, while scalable, remain fundamentally dependent on the availability and quality of labeled training data (e.g., \citealp{Dieleman2015, DominguezSanchez2018}). This presents a significant limitation in the high redshift regime, where many galaxies do not conform to the standard Hubble Sequence. If enough systems in JWST surveys have irregular or poorly understood morphologies, constructing a representative and unbiased labeled dataset becomes difficult, possibly imposing artificial structure on intrinsically diverse populations.

ZSAD provides an alternative approach that eliminates the need for labeled astronomical training data. Rather than assigning galaxies to predefined morphological categories, ZSAD reframes the problem by asking whether a given object deviates from a reference notion of normality. In this framework, galaxies with unusual morphologies are identified as anomalies, allowing the discovery of rare or complex systems without requiring prior knowledge of their appearance.

This raises a natural question: how can a model identify structural irregularities without task-specific anomaly-detection training on galaxy images? VLMs such as the Contrastive Language-Image Pre-training (CLIP; \citealt{Radford2021}) model offer a possible route. CLIP was trained on hundreds of millions of image-text pairs and learns a shared embedding space in which images and text descriptions can be directly compared. Although CLIP was not specifically designed for anomaly detection, these learned features provide a foundation for identifying structural deviations in previously unseen domains.

\subsection{AdaCLIP Framework}
\label{subsec:framework}
\subsubsection{Configurations and Model Components}
\label{subsubsec:configurations}
AdaCLIP is a ZSAD framework that adapts CLIP to anomaly detection by training additional prompt and projection layers on labeled industrial and medical imaging datasets, such as those for manufacturing defect inspection and colonic tissue abnormalities \citep{Tajbakhsh2016, Bergmann2021}. In this work, we apply available AdaCLIP checkpoints to RGB galaxy cutouts without fine-tuning or additional training. We evaluate multiple model configurations to assess AdaCLIP's sensitivity to subtle, non-standard morphological anomalies in astronomical images.

To encode visual features, AdaCLIP utilizes a Vision Transformer (ViT; \citealt{Dosovitskiy2020, Cherti2022}) backbone. A ViT processes an input image by dividing it into a sequence of non-overlapping spatial tiles called patches. Each patch is linearly projected into a feature vector and combined with a learned positional embedding that encodes its two-dimensional spatial location within the original image grid. In the ViT architectures evaluated here, the patch size is fixed at $14 \times 14$ pixels. During initial pretraining, the ViT-L/14 backbone operated on $224 \times 224$ pixel inputs, dividing images into a $16 \times 16$ grid of 256 spatial patches. The higher-resolution ViT-L/14@336 variant was pretrained on $336 \times 336$ pixel inputs, yielding a denser $24 \times 24$ grid of spatial patches. In this notation, ``L" refers to the large CLIP transformer variant \citep{Radford2021}.

AdaCLIP standardizes all input images to a uniform resolution of $518 \times 518$ pixels before processing \citep{Cao2024}. For a $14 \times 14$ pixel patch size, this generates a $37 \times 37$ grid of 1,369 spatial patches per galaxy cutout. Because this grid exceeds the native grid dimensions used during original backbone pretraining, two-dimensional bicubic interpolation is used to resample the learned positional embeddings from their pretrained grid scale ($16 \times 16$ or $24 \times 24$) to match the expanded $37 \times 37$ grid \citep{Dosovitskiy2020}. This standardized input resolution provides a consistent spatial patch grid across both architectures.

The AdaCLIP checkpoints determine the trained prompt and projection layers. AdaCLIP uses auxiliary anomaly detection datasets to optimize its hybrid learnable prompts, which combine static prompts shared across images with dynamic prompts generated from each image \citep{Cao2024}. We evaluated checkpoints trained on different dataset combinations: the All configuration (14 industrial and medical anomaly detection datasets) and two checkpoints each trained on one industrial and one medical dataset (MVTecAD \& ColonDB and VisA \& ClinicDB). Comparing the configurations allows us to assess how the choice of auxiliary training data affects ZSAD on galaxy morphologies. Table~\ref{tab:configurations} summarizes the six evaluated configurations.

\begin{deluxetable*}{ccc}
\tablecaption{Backbone architectures and AdaCLIP checkpoint configurations}
\tablehead{
\colhead{Architecture} & \colhead{AdaCLIP Checkpoint} & \colhead{Model ID}
}
\startdata
ViT-L/14 & All & \texttt{All} \\
ViT-L/14 & MVTecAD \& ColonDB & \texttt{MVTec} \\
ViT-L/14 & VisA \& ClinicDB & \texttt{VisA} \\
ViT-L/14@336 & All & \texttt{All-336} \\
ViT-L/14@336 & MVTecAD \& ColonDB & \texttt{MVTec-336} \\
ViT-L/14@336 & VisA \& ClinicDB & \texttt{VisA-336} \\
\hline
\enddata
\tablecomments{The two ViT architectures differ in the native input resolution of the CLIP image encoder (ViT-L/14: 224 $\times$ 224 pixels; ViT-L/14@336: 336 $\times$ 336 pixels), and the 14 denotes the patch size in pixels. For all configurations, galaxy cutouts are resized to $518 \times 518$ pixels before being passed to the model. The AdaCLIP checkpoints correspond to models adapted using different combinations of auxiliary anomaly-detection datasets. ``All'' denotes a checkpoint trained jointly on 14 benchmark datasets, while the remaining variants are trained on the datasets indicated in the middle column. Hereafter, we refer to each configuration by its Model ID in the rightmost column.}
\label{tab:configurations}
\end{deluxetable*}

\begin{figure*}[htb!]
\includegraphics[width=\textwidth]{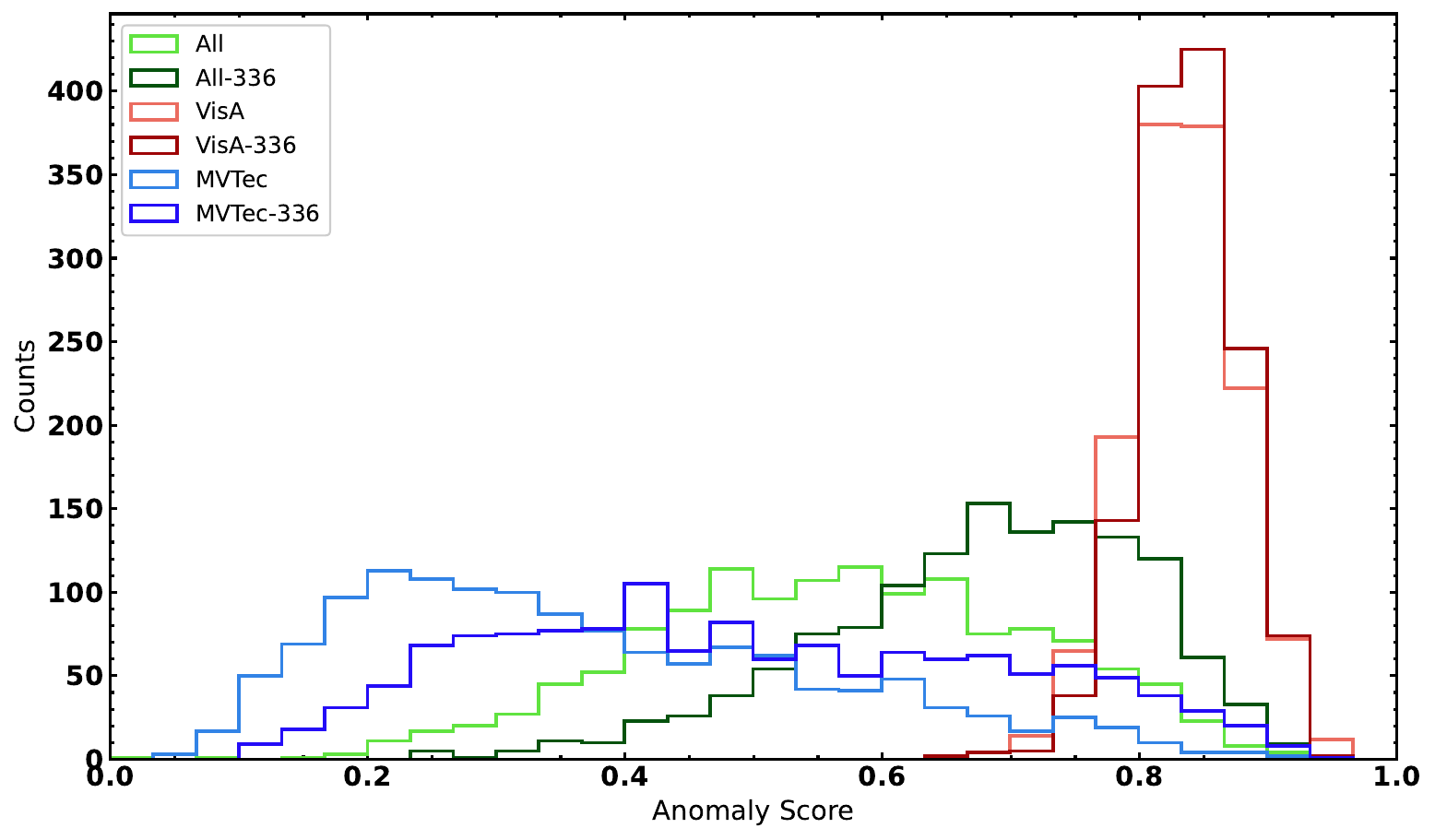}
\caption{Anomaly score distributions for 1,342 CEERS and NGDEEP objects spanning redshift $1.5 < z < 3.0$ and F277W magnitude $20 < m_{\rm{F277W}} < 24$ for the different model combinations listed in Table~\ref{tab:configurations}. Configurations using AdaCLIP checkpoints trained on the same auxiliary datasets tend to exhibit similar overall score shapes, while differences in backbone architecture produce noticeable shifts in scale and spread.}
\label{fig:score_distributions}
\end{figure*}

\subsubsection{Model Outputs}
\label{subsubsec:model_outputs}
AdaCLIP produces two outputs we consider in this work: an anomaly score and a high-dimensional embedding. The anomaly score is a single scalar in the range $[0,1]$ that measures how closely an image matches text descriptions of abnormal rather than normal states, with higher values indicating a closer match to the abnormal description. It is calculated following the method of \citet{Cao2024}, which compares an image against two sets of text descriptions. The category name (which we define to be ``galaxy") is inserted into multiple built-in normal-state and abnormal-state phrases, producing descriptions such as ``flawless galaxy" (normal) and ``damaged galaxy" (abnormal). The text encoder converts each set of descriptions into a single text vector, one representing normal states and one representing abnormal states. Separately, the ViT image encoder converts the galaxy cutout into an image vector. The anomaly score measures the relative similarity of the image vector to the normal and abnormal text vectors. Scores near 0 indicate greater similarity to the normal description, scores near 1 indicate greater similarity to the abnormal description, and scores near 0.5 indicate approximately equal similarity to both. For a fixed model configuration and input image, AdaCLIP computes the anomaly score independently of the other galaxies in the sample. Consequently, the number of galaxies evaluated does not directly affect an individual score.

In addition to the anomaly score, we extract a high-dimensional embedding for each galaxy from the AdaCLIP ViT image encoder. This embedding is the pooled global image representation: a single 768-dimensional vector that summarizes the whole cutout, taken after the encoder's final projection and before any comparison with text. Because it is computed from the image alone, it does not depend on the text prompt. We use these embeddings to compare how the different model configurations organize galaxies in embedding space.

\begin{figure*}[htb!]
\includegraphics[width=\linewidth]{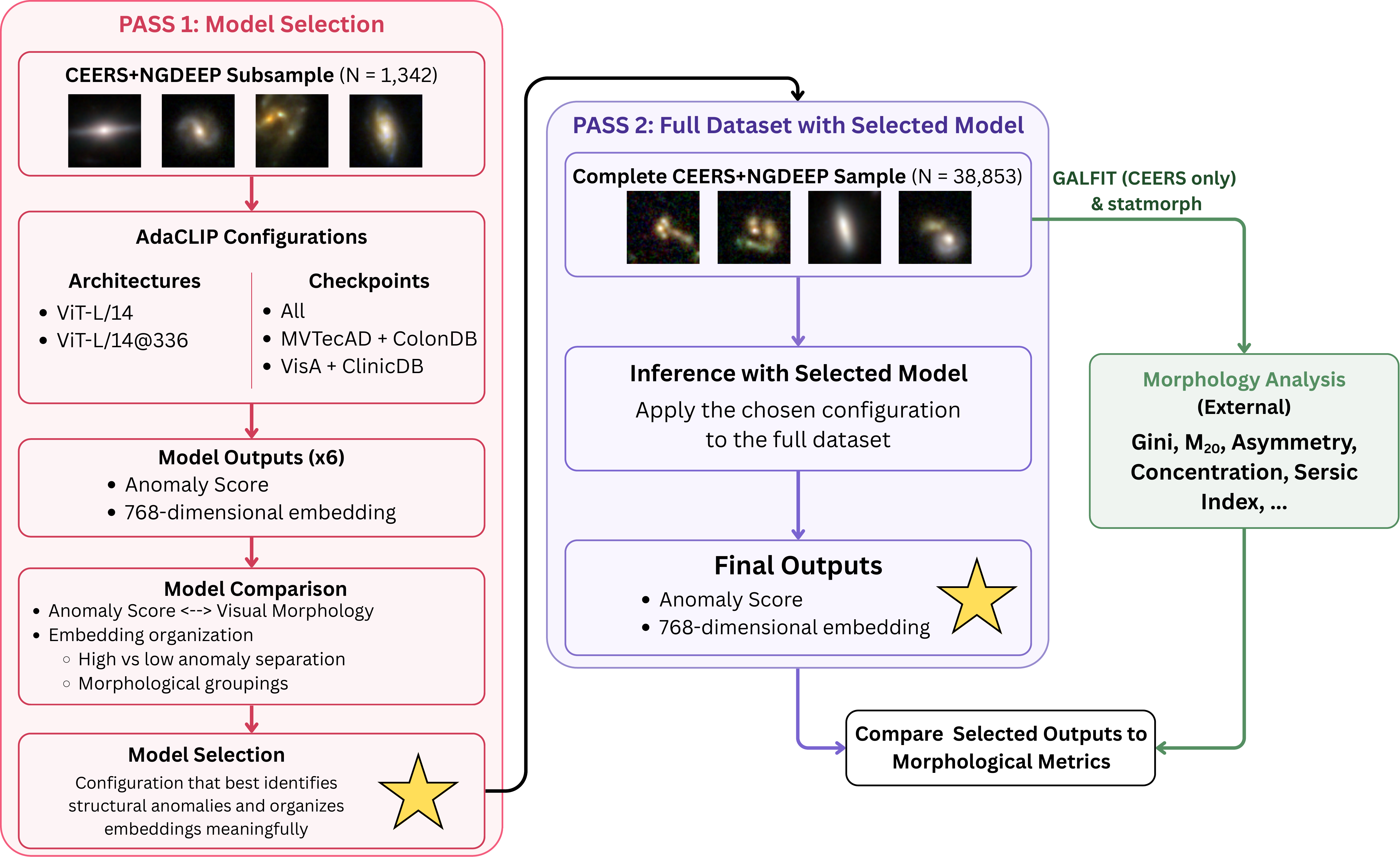}
\caption{Overview of the ZSAD pipeline. In Pass 1 (left), six AdaCLIP model configurations are tested on a subsample of 1,342 CEERS and NGDEEP galaxies, generating anomaly scores and 768-dimensional embeddings for each configuration. These outputs are compared based on their correspondence with visual morphology and the degree to which the embeddings separate high- and low-anomaly systems into coherent morphological groupings. The best-performing configuration is then selected. In Pass 2 (right), the selected model configuration is applied to the combined CEERS+NGDEEP dataset of 38,853 galaxies to generate final anomaly scores and embeddings. These outputs are subsequently compared with morphological measurements obtained using \texttt{GALFIT} (CEERS only) and \texttt{statmorph}.}
\label{fig:anomaly_detection_pipeline}
\end{figure*}

\subsection{Model Selection Criteria}
\label{subsec:model_selection}
We first evaluate the six configurations on a representative subsample of 1,342 CEERS and NGDEEP galaxies to select an optimal model setup for the entire combined catalog. Following the source selection and quality control criteria in Section~\ref{subsec:source_selection}, this benchmarking subsample isolates bright, well-detected galaxies at cosmic noon satisfying $\text{SNR}_{\text{F277W}} > 10$, $20 < m_{\rm{F277W}} < 24$, $1.5 < z < 3.0$. 

As shown in Figure~\ref{fig:score_distributions}, the resulting anomaly score distributions differ substantially between model configurations, demonstrating that both the ViT architecture and the pretraining domain influence the anomaly scores assigned to the same galaxy images. Selecting an appropriate configuration is necessary to ensure that the model used for the full sample is apt to identify structural anomalies. We compare the outputs of all six model combinations using two qualitative criteria: (1) the correspondence between anomaly scores and visually unusual galaxy morphologies and (2) the organization of galaxies with similar visual structure in the embedding space, including (a) the degree to which galaxies with high and low anomaly scores occupy distinct regions of the embedding space, and (b) the degree to which distinct morphological structures, such as spirals, ellipticals, mergers, and irregular galaxies, form recognizable groupings.

To assess whether the anomaly scores correspond to unusual galaxy morphologies, we rank all 1,342 galaxies by their anomaly scores for each model configuration and select the 20 highest-, 20 middle-, and 20 lowest-scoring objects for visual inspection. This selection is based exclusively on the anomaly score rankings, without additional visual selection criteria. We then examine the selected objects to determine which galaxy morphologies are ranked as relatively more or less anomalous. The complete set of high- and low-scoring examples for all six configurations is provided in Appendix~\ref{Appendix:high_low_comparison}. This comparison provides a qualitative assessment of whether the anomaly rankings produced by each configuration are consistent with the types of morphological irregularities targeted in this work.

We use Uniform Manifold Approximation and Projection (UMAP; \citealt{McInnes2020}) to reduce the 768-dimensional feature vectors from each model configuration into a three-dimensional latent space for interactive visual inspection. By constructing a low-dimensional manifold that prioritizes high-dimensional local neighbors, UMAP allows us to qualitatively evaluate how effectively each model groups galaxies by shared visual morphology and structural features. It is important to note that UMAP is used solely as a visualization tool; the underlying 768-dimensional embeddings remain the model output, and distances or apparent separations in the projected space should not be interpreted quantitatively.

\subsection{Pipeline Summary}
\label{subsec:pipeline}
The complete anomaly detection pipeline consists of two passes: an initial model-selection pass followed by application of the selected configuration to the full galaxy sample. The workflow is summarized in Figure~\ref{fig:anomaly_detection_pipeline}. In Pass 1, all six AdaCLIP configurations are applied to the subsample of CEERS and NGDEEP galaxies. The resulting anomaly scores and embeddings are then evaluated using the model-selection criteria described in Section~\ref{subsec:model_selection} to identify the configuration most appropriate for detecting structural anomalies. In Pass 2, the selected AdaCLIP configuration is applied to the full CEERS+NGDEEP sample. The resulting anomaly catalog is then used in conjunction with independent morphological measurements in the analyses presented in Section~\ref{sec:results}.

\section{Measures of Galaxy Morphology}
\label{sec:galaxy_morphology}
Morphological measurements are commonly divided into parametric and nonparametric approaches, which characterize galaxy structure using analytic light-profile models and model-independent statistics, respectively. We obtain parametric measurements for CEERS galaxies with available \texttt{GALFIT} \citep{Peng2002} fits from \citet{McGrath2026} and compute nonparametric measurements for the CEERS+NGDEEP sample using \texttt{statmorph} \citep{RodriguezGomez2019}. The following section introduces the morphology measurements used throughout this work.

\subsection{Parametric Measurements}
\label{subsec:parametric}
The most commonly used parametric approach is based on the Sérsic profile \citep{Sérsic1963}, which provides a compact description of galaxy structure through parameters such as the Sérsic index, $n$, and effective radius, $R_{e}$, defined here as the semimajor axis of the elliptical region enclosing half of the total modeled galaxy light. The Sérsic index quantifies the concentration of a galaxy's light profile, with disk-dominated systems generally exhibiting lower values ($n\sim1$) and bulge-dominated systems exhibiting higher values ($n\sim4$).

We obtain positional and morphological measurements from the \citet{McGrath2026} CEERS \texttt{GALFIT} catalog. The morphology catalog was generated using an independent source detection and photometry pipeline and therefore does not have a one-to-one correspondence with the official CEERS photometric catalog. \citet{McGrath2026} perform nearest-neighbor matches between their CEERS objects and the official CEERS catalog; however, differences in source detection algorithms can lead to discrepancies in sky positions, particularly for large galaxies that may be split into multiple components or merged into a single detection. Because the right ascension and declination values differ slightly between the catalogs, we perform our own nearest-neighbor cross-match between the 34,445 CEERS galaxies that pass our quality criteria (see Table~\ref{tab:data_reduction}) and the CEERS morphology catalog, using a matching radius of $1\farcs0$.

For the cross-matched galaxies, the CEERS morphology catalog provides \texttt{GALFIT} measurements in the same NIRCam filters used in this work. To ensure consistent comparison across redshift, we adopt a filter-selection procedure designed to trace approximately the same rest-frame optical emission for all galaxies. We assume a reference rest-frame wavelength of $\lambda_{\mathrm{rest}} = 0.5$ $\mu$m and compute the corresponding observed wavelength as
\begin{equation}
\lambda_{\mathrm{obs}} = \lambda_{\mathrm{rest}}(1+z),
\label{eq:restframe}
\end{equation}
where $z$ is the photometric redshift of each galaxy. We select the NIRCam filter whose central wavelength is closest to $\lambda_{\mathrm{obs}}$ and adopt the corresponding \texttt{GALFIT}-derived structural parameters. This per-object filter assignment reduces wavelength-dependent morphological effects and provides a consistent basis for comparison across the entire redshift range.

After cross-matching objects and assigning NIRCam filters, reliable parametric morphology measurements are available for 8,272 CEERS galaxies in our sample. The distributions and relationships between the \texttt{GALFIT} measurements and the anomaly detection results are examined in Section~\ref{subsec:parametric_results}.

\subsection{Nonparametric Measurements}
\label{subsec:nonparametric}
We measure nonparametric morphologies using \texttt{statmorph} \citep{RodriguezGomez2019}. For each of the 38,853 galaxies in the CEERS+NGDEEP dataset, we first identify the NIRCam filter whose central wavelength most closely matches the galaxy's rest-frame optical emission, following the filter selection procedure described in Section~\ref{subsec:parametric} (Equation~\ref{eq:restframe}). We then run \texttt{statmorph} only on the selected filter, using the corresponding science image cutout, segmentation map, error map, and empirical PSF \citep{Cox2026}. The segmentation and error-map cutouts are matched to the spatial dimensions of the science image cutout.

\texttt{statmorph} measures a wide range of morphology statistics commonly used in extragalactic astronomy. The quantities used in this work are concentration, asymmetry, and smoothness ($CAS$) \citep{Bershady2000, Conselice2000, Conselice2003}, as well as the Gini coefficient $(G)$, which measures how flux is distributed among a galaxy's pixels, with higher values indicating that a larger fraction of the light is concentrated in fewer pixels, and the second moment of the region of the galaxy containing $20\%$ of the total flux $(M_{20})$ \citep{Abraham2003, Lotz2004}.

The asymmetry statistic measures deviations from $180^\circ$ rotational symmetry in a galaxy's light distribution after correcting for background contributions, with larger values indicating greater asymmetry \citep{Conselice2003, RodriguezGomez2019}. We apply quality cuts to the resulting \texttt{statmorph} measurements, removing objects with quality flags greater than $1$ or extreme negative asymmetry values $(A < -1.5)$. Although background corrections can produce negative asymmetry values, such strongly negative values are treated as unreliable rather than physically meaningful measurements of galaxy structure. In total, 33,976 galaxies satisfy these criteria and are retained for the subsequent analysis. The excluded galaxies exhibit a broadly similar redshift distribution to the retained sample but are relatively more prevalent at brighter F277W magnitudes ($m_{\mathrm{F277W}}\lesssim24$).

\section{Results}
\label{sec:results}
Following the model-configuration comparison (Section~\ref{subsec:model_selection_results}, we examine the relationships between anomaly scores and galaxy photometry (Section~\ref{subsec:photometry_results}), parametric morphological measurements from \texttt{GALFIT} (Section~\ref{subsec:parametric_results}), and nonparametric measurements from \texttt{statmorph} (Section~\ref{subsec:nonparametric_results}). For each quantity, we present its distribution, the median anomaly score within bins of that measurement, and the corresponding UMAP embedding colored by the same quantity to examine how these properties are organized in the learned representation space.

\begin{figure*}[htb!]
\includegraphics[width=\linewidth]{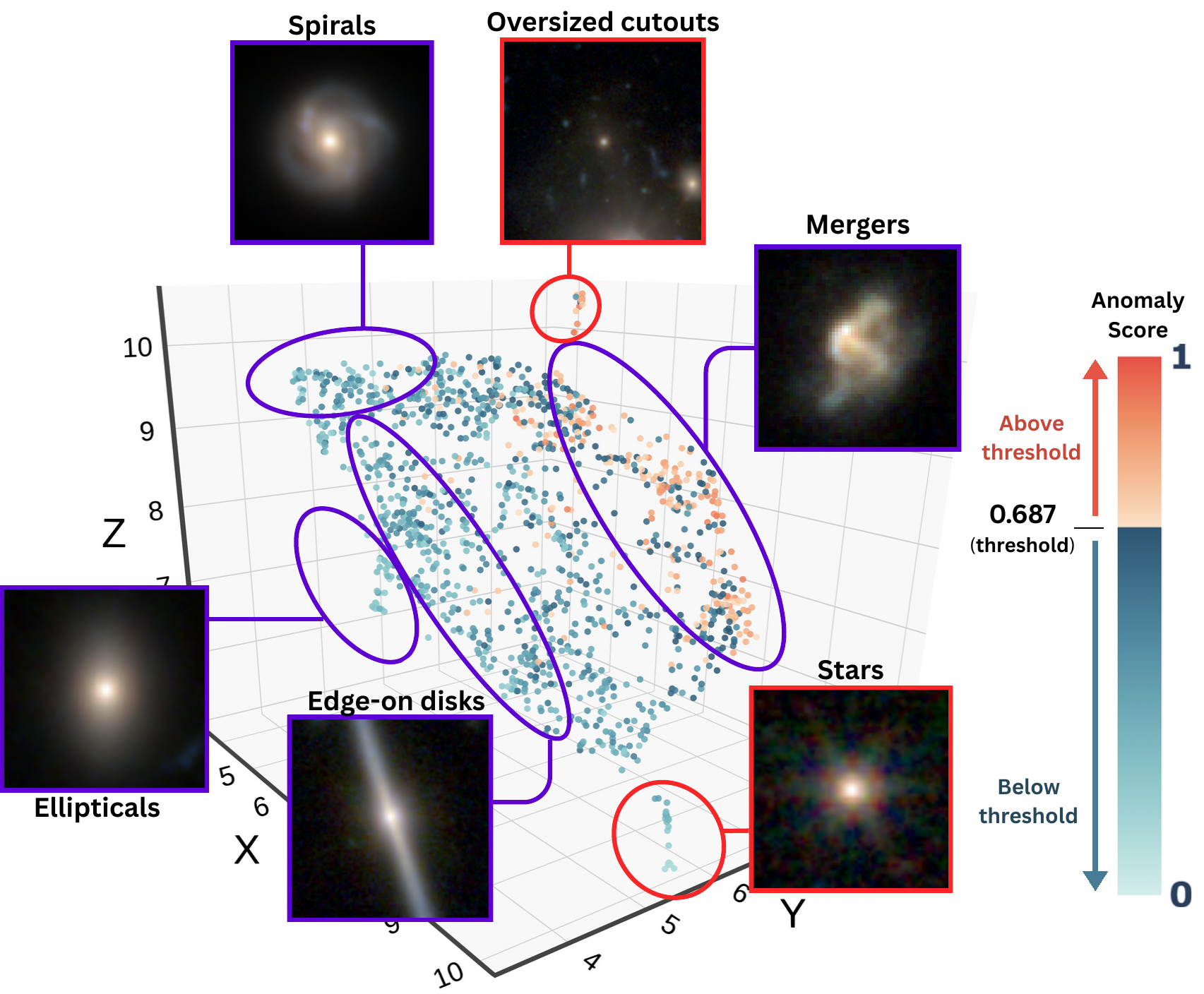}
\caption{UMAP projection of the 768-dimensional AdaCLIP embeddings for the 1,342 CEERS and NGDEEP objects spanning $1.5 < z < 3$ and $20 < m_{\rm{F277W}} < 24$, using the \texttt{MVTec-336} model configuration, rotated to best show the resulting groupings. Points are colored by anomaly score, with a cutoff at the 80th percentile of the score distribution (0.687) separating most visually disturbed objects from undisturbed ones; high-anomaly points are concentrated within the mergers/irregulars grouping. Selected morphological categories are highlighted: elliptical and spiral morphologies occupy different regions of the embedding space despite exhibiting similar anomaly scores; mergers and irregular systems appear clustered together; edge-on disks also occupy a distinct region of the latent space. The model also separates contaminants such as stars and oversized cutouts from the normal population.}
\label{fig:UMAP_CEERS}
\end{figure*}

\subsection{AdaCLIP Model Selection}
\label{subsec:model_selection_results}
Figures~\ref{fig:low_high_comparison_all}--\ref{fig:low_high_comparison_visa} show the 20 highest-, middle-, and lowest-scoring galaxies for each model configuration (see Appendix~\ref{Appendix:high_low_comparison}). Across the six configurations, objects with the highest anomaly scores generally exhibit disturbed or uncommon morphologies, including mergers, tidal features, and asymmetric light distributions. The middle-scoring galaxies exhibit a broader mixture of morphologies, including relatively regular disks, centrally concentrated systems, and galaxies with more noticeable structural irregularities. These intermediate examples illustrate that the anomaly rankings do not necessarily correspond to sharply separated morphological classes. In contrast, models vary in the morphologies to which they assign low scores. In \texttt{All}, the lowest-scoring objects are predominantly centrally concentrated with diffuse outer halos, whereas \texttt{All-336} assigns edge-on galaxies the lowest scores. \texttt{MVTec} and \texttt{MVTec-336} both assign low scores to centrally concentrated and edge-on systems. The All and MVTecAD configurations therefore generally produce a desirable ranking in which visually disturbed systems occupy the high-score end, while relatively regular systems occupy the low-score end.

The \texttt{VisA} and \texttt{VisA-336} configurations similarly assign low scores primarily to centrally concentrated or edge-on systems. However, both configurations also place some galaxies with visible tidal features among the lowest-scoring objects. Because these features represent the types of morphological irregularities we aim to identify, their presence among the lowest-scoring systems indicates a less consistent correspondence between anomaly score and visual morphology for these configurations.

The three-dimensional UMAP projections for all six models are shown in Appendix~\ref{Appendix:embedding_space}. To establish a practical threshold for identifying structurally anomalous systems, we visually inspected the CEERS subsample across the range of anomaly scores for each model configuration. We find that galaxies at or above the 80th percentile of the respective anomaly score distribution generally exhibit visually distinct or structurally anomalous morphologies, motivating its use as the boundary for the high-anomaly population. As such, each projection is colored by anomaly score, with the high-anomaly population defined relative to the 80th percentile of the score distribution for that configuration. Using a percentile-based division rather than a fixed threshold value accounts for the different score scales produced by the six models.

\begin{figure*}[htb!]
\hspace{0.6in}
\includegraphics[width=0.85\textwidth]{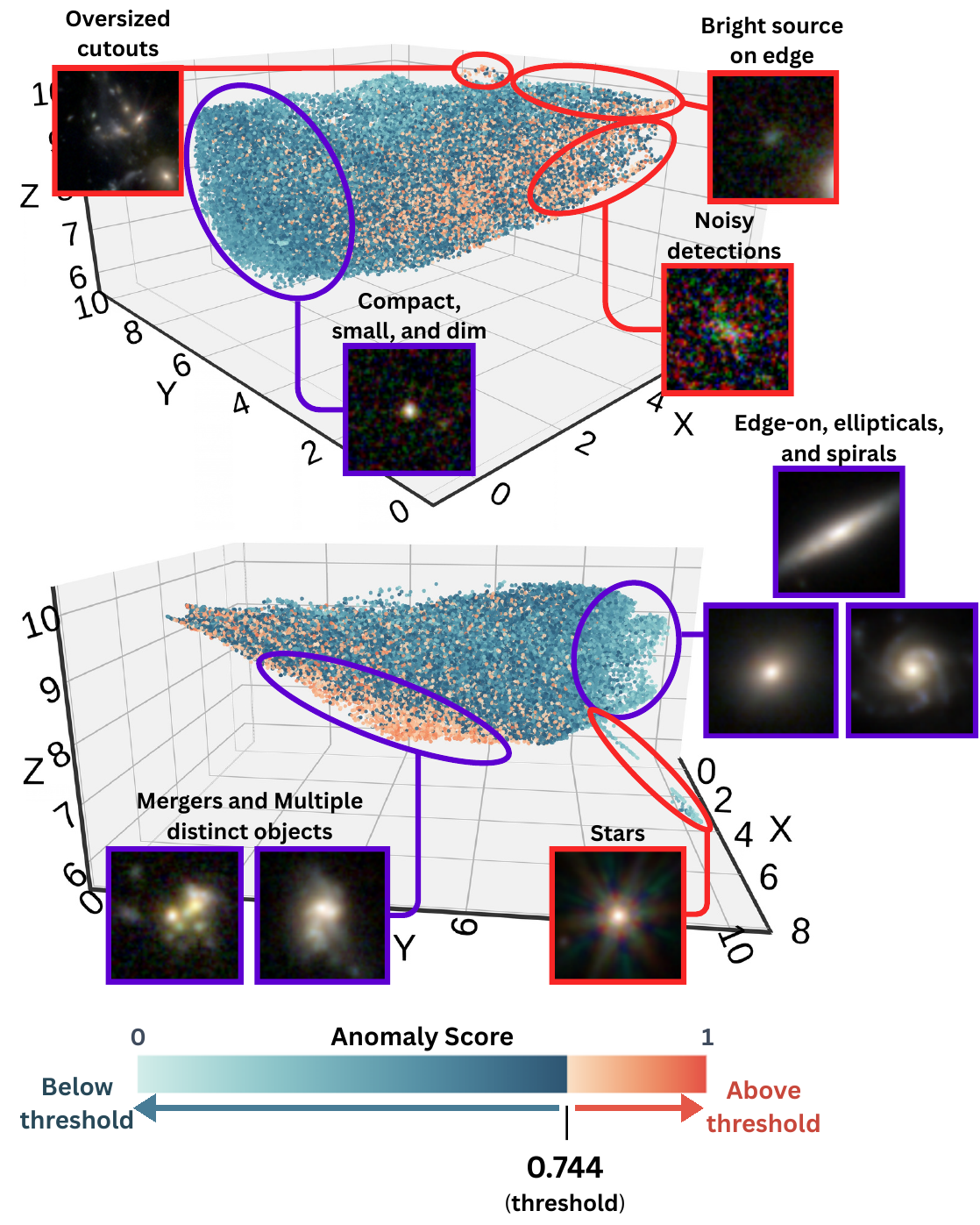}
\caption{UMAP projection of the 768-dimensional AdaCLIP embeddings for the 38,853 galaxies in the combined CEERS+NGDEEP dataset. The top and bottom panels are rotated to best show the highlighted groups. Points are colored by anomaly score, with a cutoff at the 80th percentile of the score distribution (0.744) separating most visually disturbed objects from undisturbed ones; high-anomaly points are concentrated within the mergers/irregulars grouping. Selected morphological categories are highlighted: edge-on, elliptical, and spiral morphologies occupy a similar region of the embedding space, while mergers, irregulars, and cutouts with multiple distinct objects occupy a separate region. Contaminants such as stars, oversized cutouts, noisy detections, and cutouts with bright sources on the edge are highlighted as well.}
\label{fig:UMAP_CEERS+NGDEEP}
\end{figure*}

\begin{figure*}[htb!]
\includegraphics[width=\textwidth]{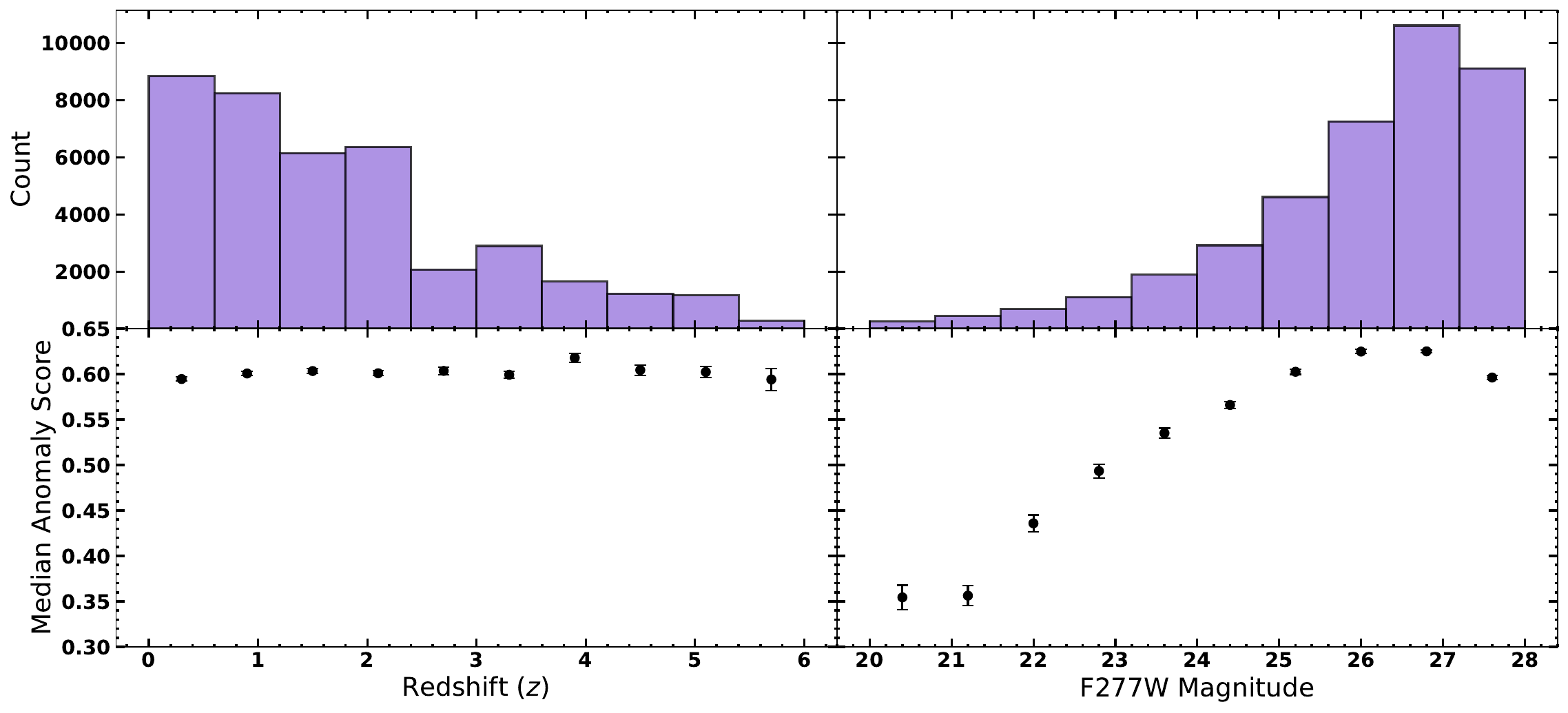}
\caption{Top: redshift (left) and F277W magnitude (right) distributions of the final sample of 38,853 CEERS and NGDEEP galaxies. Bottom: median anomaly score within each bin. The error bars represent the standard error of the median.}
\label{fig:photometry_results}
\end{figure*}

\begin{figure*}[htb!]
\includegraphics[width=\textwidth]{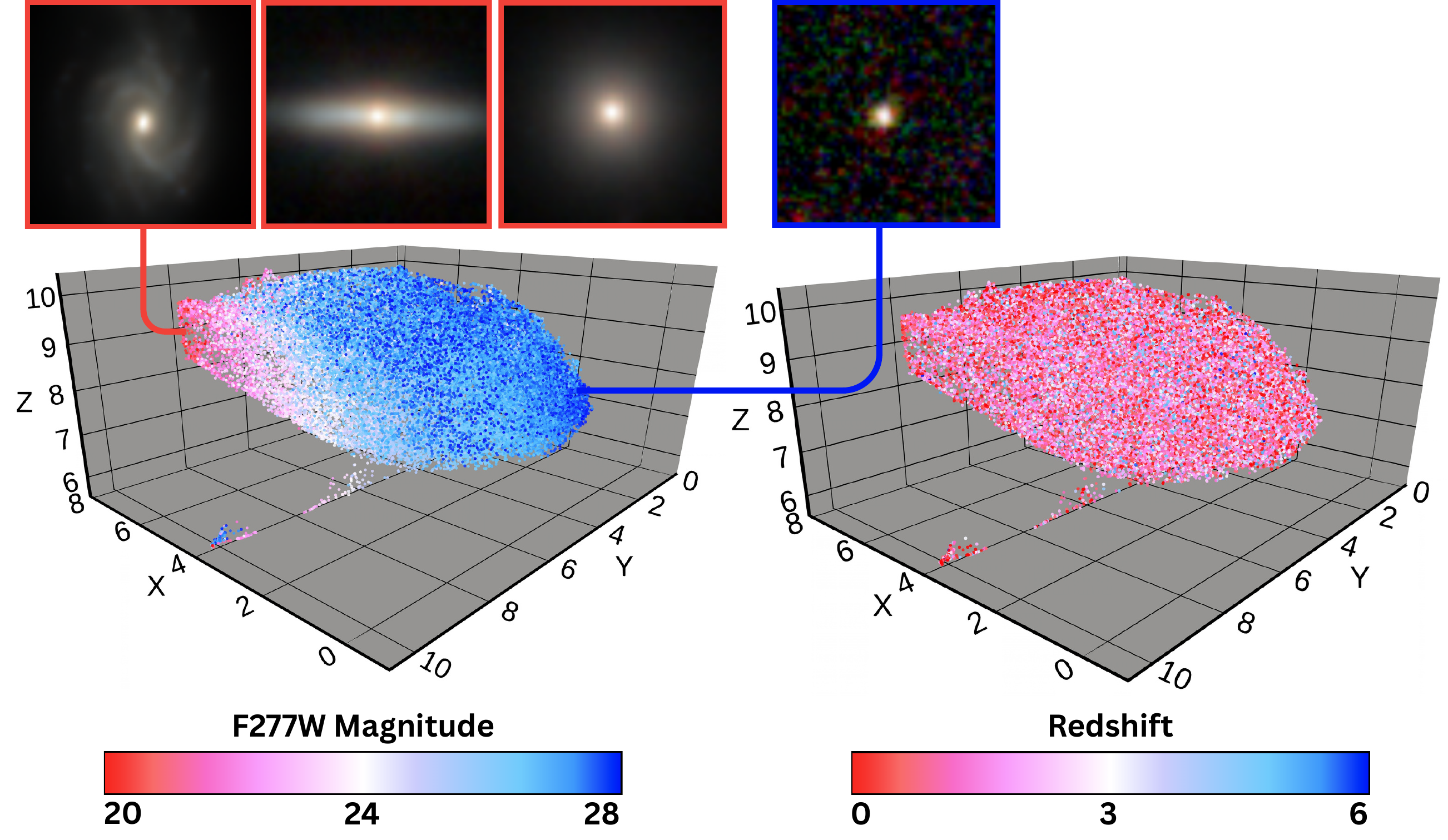}
\caption{Three-dimensional UMAP projection of the full sample of 38,853 CEERS and NGDEEP galaxies, colored by F277W magnitude (left) and redshift (right). Each panel shows the same embedding space with an independent colorbar. Red-outlined cutouts above the top panel are representative galaxies from the bright end of the embedding, and the blue-outlined cutout is a representative galaxy from the faint end. No cutouts are shown for redshift.}
\label{fig:photometric_embeddings}
\end{figure*}

In each model configuration, galaxies with similar morphologies tend to occupy nearby regions of the embedding space. However, the distribution of high-anomaly objects differs between models. The four \texttt{All} and \texttt{VisA} projections do not show a clear concentration of high-anomaly objects in a distinct region of the embedding space. In contrast, both \texttt{MVTec} configurations show localized regions containing higher concentrations of high-anomaly objects, consistent with the visual morphologies of galaxies in these regions. The \texttt{MVTec} projection also shows some separation between broad morphological classes, particularly spirals, ellipticals, and mergers, although these groups are relatively diffuse and do not form sharply defined regions. Of the two MVTecAD configurations, the \texttt{MVTec-336} projection provides the clearest organization of high-anomaly systems relative to the surrounding population and also produces stronger separation between broad morphological classes.

As shown in Figure~\ref{fig:UMAP_CEERS}, the \texttt{MVTec-336} model isolates high-anomaly objects into a primary region of the embedding space that is separate from the surrounding lower-scoring population. Within the lower-anomaly population, galaxies with similar anomaly scores but different morphologies, such as ellipticals and spirals, also occupy distinct regions of the embedding space. This organization suggests that the model captures structural differences between galaxies beyond those reflected in the scalar anomaly score alone. The model also groups contaminants such as stars and oversized cutouts into distinct anomalous peripheral clusters. This indicates that the \texttt{MVTec-336} space simultaneously captures meaningful morphological variations and isolates non-galaxy artifacts. Because it satisfies both qualitative evaluation criteria outlined in Section~\ref{subsec:model_selection}, we select the \texttt{MVTec-336} model configuration for application to the full dataset.

\begin{figure*}[htb!]
\includegraphics[width=\textwidth]{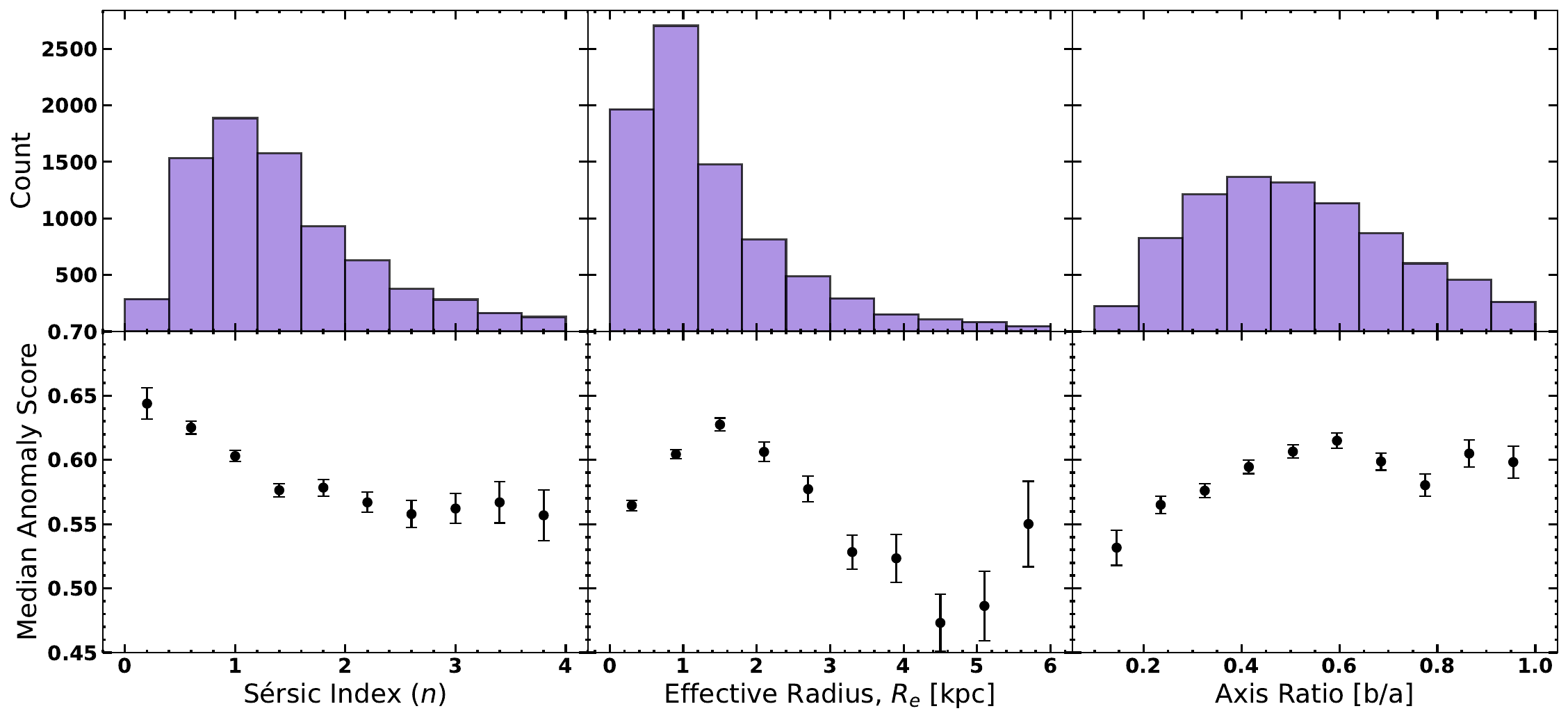}
\caption{Top: Sérsic index (left), size (center), and axis ratio (right) distributions of the 8,272 CEERS galaxies with reliable \texttt{GALFIT} fits from \citet{McGrath2026}. Bottom: median anomaly score within each bin.}
\label{fig:parametric_results}
\end{figure*}

We extend the anomaly detection pipeline to the combined CEERS+NGDEEP dataset of 38,853 galaxies. The embeddings are projected into the three-dimensional UMAP space shown in Figure~\ref{fig:UMAP_CEERS+NGDEEP}. Although the combined sample is more diverse, the model continues to organize objects according to their anomaly scores, with higher-scoring objects occupying a distinct region of the embedding space. Examination of the embedding space from multiple viewing angles also reveals several morphological regions. We find that most galaxies in the combined dataset are visually compact, small, and dim. Due to the sheer number of these galaxies, the separation of spiral, elliptical, and edge-on systems is not as strong or as apparent as in Figure~\ref{fig:UMAP_CEERS}. Although these galaxies inhabit a single region of the embedding space, they do show separation amongst themselves. Several additional groupings of artifacts and contaminants emerge, including stars; noisy detections; oversized cutouts; and bright sources on the edge of cutouts. Most notably, a distinct group of high-anomaly objects persists, dominated by visually disturbed galaxies and systems with multiple components, many of which exhibit features consistent with mergers. This supports the model selection and demonstrates that the \texttt{MVTec-336} configuration reliably isolates potentially interacting systems from both normal galaxy morphologies and contaminants at scale.

\subsection{Photometric Properties and Anomaly Score}
\label{subsec:photometry_results}
Figure~\ref{fig:photometry_results} presents the distributions of redshift and $m_{\rm{F277W}}$ for the 38,853 galaxies in the combined sample, together with the median AdaCLIP anomaly score within the same bins. The redshift distribution peaks in the lowest bin and declines toward higher redshift. The median anomaly score is essentially flat across the full range of $0<z<6$. $m_{\rm{F277W}}$ shows a much stronger relationship, rising toward fainter values and falling slightly in the highest bin. The median anomaly score is lowest for the brightest galaxies ($\sim 0.36$ at $m_{\rm{F277W}} < 21$) and increases with $m_{\rm{F277W}}$ to $\sim 0.60$, then declines slightly in the faintest bin.

\begin{figure*}[htb!]
\includegraphics[width=\textwidth]{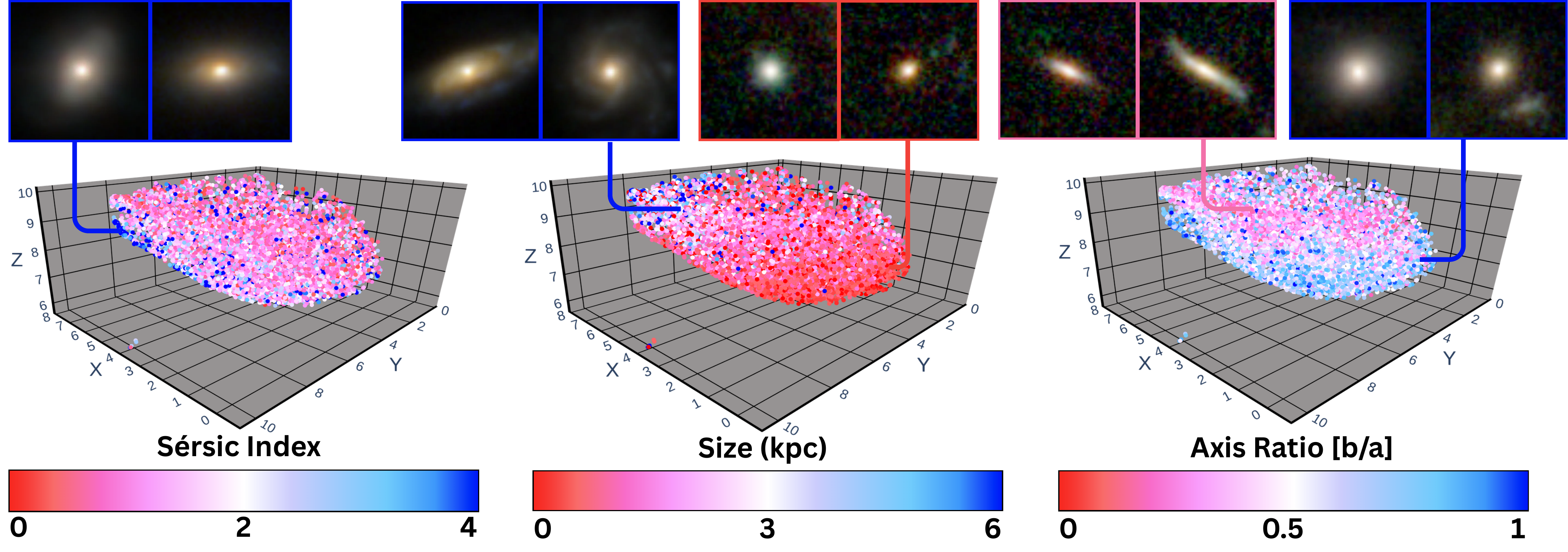}
\caption{Three-dimensional UMAP projection of CEERS galaxies with \texttt{GALFIT} measurements from \cite{McGrath2026}, colored by Sérsic index (left), effective radius in kpc (center), and axis ratio (right). Each panel shows the same embedding with an independent colorbar, rotated to best show any groupings or clusters that emerge. Cutouts above each panel are representative galaxies from selected regions of the embedding. High-$n$ galaxies show a weak concentration toward one edge of the space, while effective radius and axis ratio vary more smoothly across the projection.}
\label{fig:parametric_embeddings}
\end{figure*}

Figure~\ref{fig:photometric_embeddings} shows the three-dimensional UMAP projection colored by $m_{\rm{F277W}}$ and redshift. Apparent magnitude exhibits a clear spatial organization in the embedding. Galaxies are separated by $m_{\rm{F277W}}$ across the projection, forming a smooth, continuous gradient from the brightest systems at one edge to the faintest systems at the opposite edge, with no abrupt boundaries between intermediate values. The bright galaxies occupy a distinct region of the space, and the cutouts drawn from this region show a wide range of morphologies, including a face-on spiral, an edge-on disk, and a smooth elliptical. The faint end contains the bulk of the sample, and the cutout drawn from this region shows a small, compact source. Redshift shows no comparable organization. Galaxies spanning the full range of $z$ are interspersed throughout the projection, with no evidence of spatial grouping or gradient. This is consistent with the absence of a trend between redshift and anomaly score in Figure~\ref{fig:photometry_results}. We therefore show no cutouts in the right panel.

\begin{figure*}[htb!]
\includegraphics[width=\textwidth]{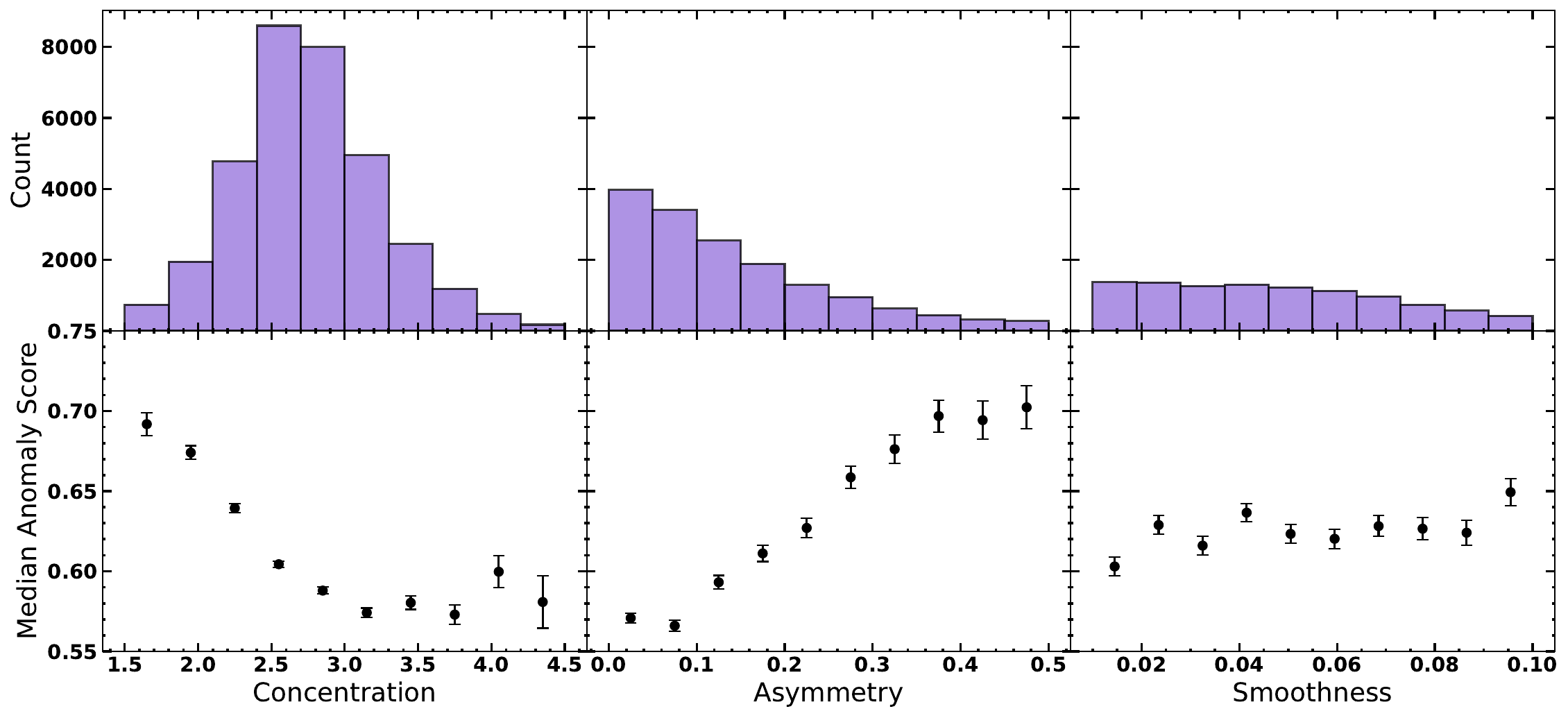}
\caption{Top: concentration (left), asymmetry (center), and smoothness (right) distributions of the 33,976 CEERS and NGDEEP galaxies with reliable \texttt{statmorph} measurements. Bottom: median anomaly score within each bin.}
\label{fig:CAS_results}
\end{figure*}

\subsection{Parametric Morphology and Anomaly Score}
\label{subsec:parametric_results}
Figure~\ref{fig:parametric_results} presents the distributions of Sérsic index (where $n = 0.5$ corresponds to a Gaussian profile, $n = 1$ corresponds to an exponential profile, and $n=4$ to a de Vaucouleurs profile), size (effective radii, $R_{e}$), and axis ratio ($q$) for the 8,272 CEERS galaxies with reliable \texttt{GALFIT} measurements. The Sérsic index distribution peaks near $n \sim 1$, consistent with a disk-dominated population, with a broad tail extending toward higher $n$. The corresponding median anomaly score generally decreases with increasing Sérsic index before flattening at $n \gtrsim 2.5$. Size exhibits a different trend. The effective radius distribution peaks at small sizes, with most galaxies falling below $R_{e} \sim 2$ kpc and declining toward larger systems. The median anomaly score increases from the smallest size bins to a peak near $R_{e} \sim 1.5$ kpc, then declines toward larger radii. The axis ratio distribution is wider and more evenly populated. The median anomaly score increases from low to intermediate axis ratios, peaks near $q \sim 0.6$, and flattens at the highest values.

\begin{figure*}[htb!]
\includegraphics[width=\textwidth]{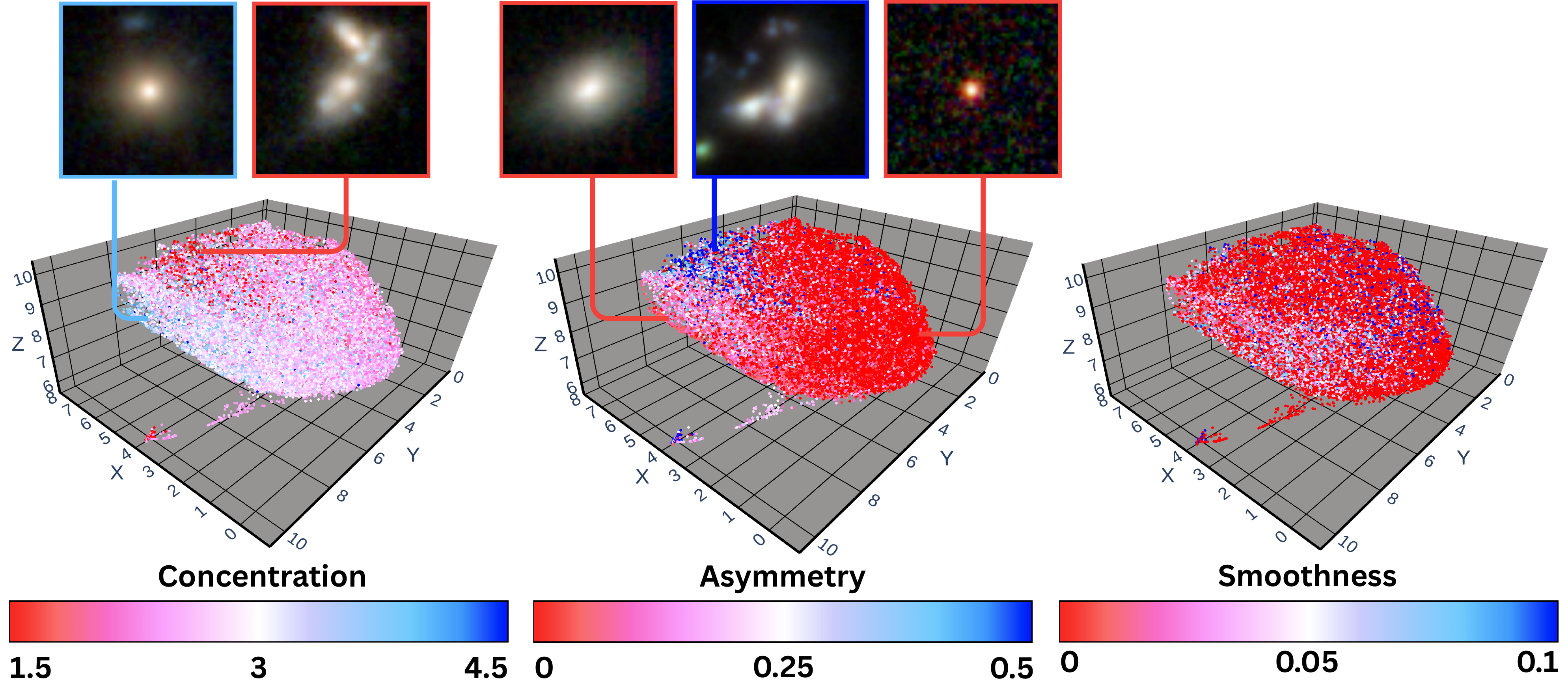}
\caption{Three-dimensional UMAP projections of galaxies with reliable \texttt{statmorph} measurements, colored by concentration (left), asymmetry (center), and smoothness (right). Each panel shows the same embedding space with an independent colorbar, rotated to best show any groupings or clusters that emerge. Cutouts above the concentration and asymmetry panels are representative galaxies from selected regions of the embedding. No cutouts are shown for smoothness.}
\label{fig:nonparametric_embeddings}
\end{figure*}

\begin{figure*}[htb!]
\includegraphics[width=\textwidth]{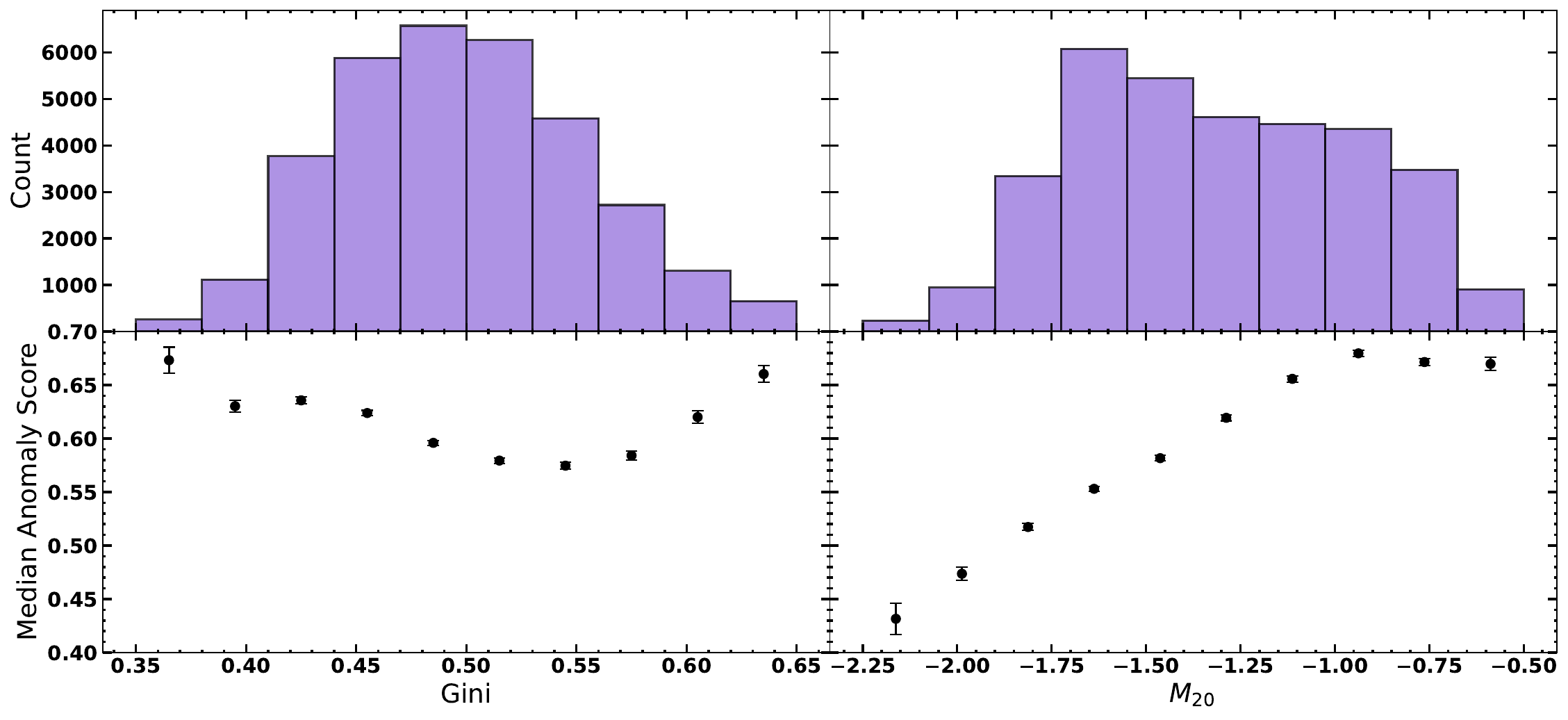}
\caption{Top: Gini (left) and $M_{20}$ (right) distributions of the 33,976 CEERS and NGDEEP galaxies with reliable \texttt{statmorph} measurements. Bottom: median anomaly score within each bin.}
\label{fig:Gini-M20_results}
\end{figure*}

Figure~\ref{fig:parametric_embeddings} shows the three-dimensional UMAP projection of CEERS galaxies with \texttt{GALFIT} measurements, colored by Sérsic index, effective radius, and axis ratio. Sérsic index reveals moderate spatial segregation: galaxies with low and high $n$ are interspersed throughout much of the projection, although high-$n$ systems are somewhat concentrated toward one edge of the space. Effective radius shows clearer structure, with smaller galaxies concentrated on one side of the projection and larger galaxies on the opposite side, forming a continuous gradient across the space. This distribution qualitatively resembles the gradient in $m_{\mathrm{F277W}}$ shown in Figure~\ref{fig:photometric_embeddings}, with larger galaxies tending to occupy regions associated with brighter sources. Axis ratio shows a comparable pattern: elongated and round systems occupy visually distinct regions, with intermediate ($q \sim 0.5$) systems positioned between them. The representative cutouts for effective radius and axis ratio reflect these trends, with galaxies at opposite ends of each gradient appearing at opposite ends of the projection.

\subsection{Nonparametric Morphology and Anomaly Score}
\label{subsec:nonparametric_results}
Distributions of concentration ($C$), asymmetry ($A$), and smoothness ($S$) for the sample of 33,976 galaxies with reliable \texttt{statmorph} measurements are shown in Figure~\ref{fig:CAS_results}. We exclude galaxies with negative asymmetry values from the asymmetry panel and restrict the smoothness panel to $0.01<S<0.1$, omitting potentially unreliable measurements at $S=0$. The concentration distribution peaks near $C \sim 2.7$, with the median anomaly score generally decreasing from approximately 0.69 at low concentrations to 0.57 near $C \sim 3.5$. The asymmetry distribution is concentrated toward low values, with the number of galaxies declining gradually as $A$ increases. In contrast to concentration, the median anomaly score increases steadily with asymmetry, from approximately 0.57 in the lowest bins to 0.70 at the highest values. The smoothness distribution is relatively flat at lower values and gradually declines toward higher $S$. The median anomaly score varies only weakly with smoothness, remaining near $\sim 0.63$ across most bins, with a slightly lower value of $\sim 0.60$ in the first bin and $\sim 0.65$ in the highest-$S$ bin.

\begin{figure*}[htb!]
\centering
\includegraphics[width=\textwidth]{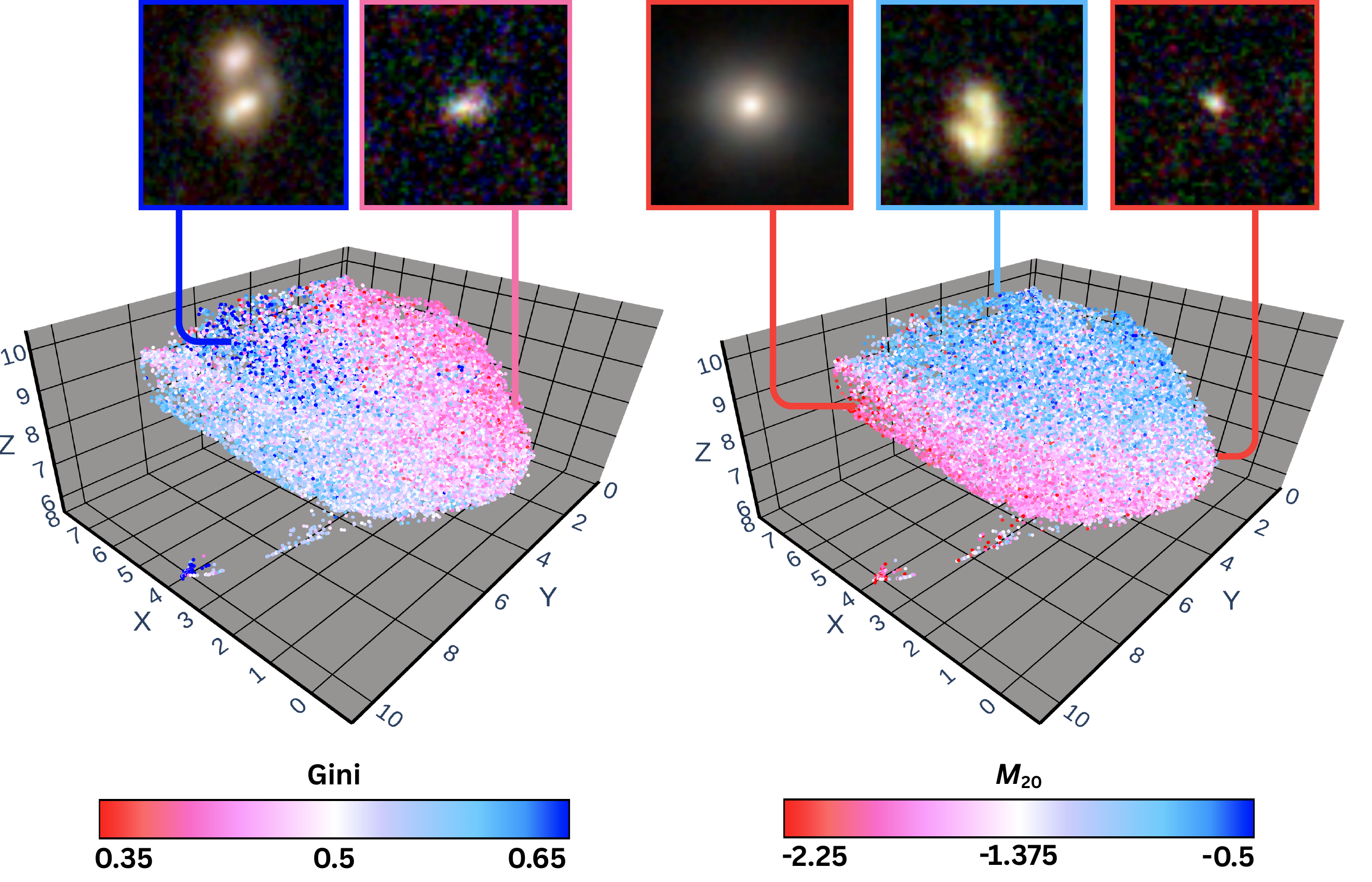}
\caption{Three-dimensional UMAP projection of galaxies with \texttt{statmorph} measurements, colored by Gini coefficient (left) and $M_{20}$ (right). Each panel shows the same embedding space with an independent colorbar, rotated to best show any groupings or clusters that emerge. Cutouts above each panel are representative galaxies from selected regions of the embedding.}
\label{fig:Gini_embeddings}
\end{figure*}

Figure~\ref{fig:nonparametric_embeddings} shows three-dimensional UMAP projections of the same sample, colored by concentration, asymmetry, and smoothness. Concentration exhibits some spatial organization, with a localized group of high-$C$ galaxies containing bright, centrally concentrated systems, while lower-concentration galaxies appear in another region that includes disturbed, multi-component systems. Galaxies with low-to-intermediate concentration values are also distributed broadly throughout the embedding. Asymmetry reveals a more distinct separation, with the highest-$A$ galaxies concentrated in a relatively confined region containing merging and interacting systems. Low-$A$ galaxies, however, occupy multiple regions of the embedding, including one containing relatively bright, well-defined galaxies and another dominated by faint, compact sources with visible background noise. In contrast, smoothness shows little spatial organization, with galaxies spanning a range of $S$ values distributed throughout the projection without clearly defined groupings.

Figure~\ref{fig:Gini-M20_results} presents the distributions of the Gini coefficient ($G$) and $M_{20}$ for the same sample. The Gini distribution peaks near $G \sim 0.5$, declining toward both lower and higher values. Its relationship with anomaly score is nonmonotonic: the median score is relatively high in the lowest-$G$ bins, decreases toward a minimum near $G \sim 0.55$, and rises again at the highest Gini values. The $M_{20}$ distribution is broader, peaking near $M_{20} \sim -1.8$ before declining gradually toward less negative values, corresponding with a greater contribution from spatially offset bright regions. Unlike Gini, the median anomaly score increases consistently as $M_{20}$ becomes less negative, from approximately 0.43 in the most negative bins to 0.67 near $M_{20} \sim -1.0$, after which it remains relatively constant.

The corresponding UMAP projections in Figure~\ref{fig:Gini_embeddings} reveal spatial variations in both measurements. Coloring by Gini coefficient highlights a localized group of high-$G$ galaxies at one end of the embedding, including systems with multiple bright components, while low-$G$ galaxies are more prevalent toward the opposite end. Intermediate values remain distributed throughout much of the projection. The $M_{20}$ embedding shows a different pattern, with the lowest values concentrated in two spatially separated regions. Representative cutouts from these regions include both a bright, centrally concentrated galaxy and a faint, compact source, illustrating the different visual appearances associated with low $M_{20}$. Galaxies with intermediate-to-high $M_{20}$ values are distributed more broadly across the embedding, including a region containing bright, irregular, and multicomponent systems.

\section{Discussion}
\label{sec:discussion}
\subsection{Morphological Information Decoded by AdaCLIP}
\label{subsec:AdaCLIP_info}
Although AdaCLIP was developed as a general-purpose anomaly detector, the choice of auxiliary anomaly-detection dataset strongly influences the quality of the resulting anomaly scores and embedding space. Among the six model configurations explored in Section~\ref{subsec:model_selection}, the configurations adapted using ColonDB and MVTecAD datasets \citep{Tajbakhsh2016, Bergmann2021} consistently produce the most visually sensible anomaly scores and the clearest morphological organization within the latent space. While these datasets are unrelated to astronomy, they share a common learning objective: isolating structural abnormalities from an otherwise coherent background. In MVTecAD, these abnormalities comprise defects like scratches or cracks in manufactured materials, whereas ColonDB anomalies represent polyp structures in normal tissue. Likewise, the highest anomaly scores in our galaxy sample correspond to disturbed morphologies, multiple components, asymmetric light distributions, diffuse emission, or other structural departures from a smooth baseline. The visual similarity between localized defects in industrial inspection and local irregularities in galaxy cutouts may help explain why the learned anomaly-detection representations transfer to astronomical imaging.

The choice of vision transformer (ViT) backbone resolution further modulates this feature extraction. Both the \texttt{ViT-L/14} and \texttt{ViT-L/14@336} variants process inputs resampled to a $518 \times 518$ pixel grid ($37 \times 37$ spatial patches), but their positional embeddings were originally learned on different grid resolutions prior to bicubic interpolation. The denser pretraining grid of \texttt{ViT-L/14@336} may provide more detailed learned positional representations, potentially influencing its sensitivity to localized structural features, such as faint tidal tails, compact companion nuclei, and clumpy star-forming regions. This may contribute to the clearer morphological organization observed for the \texttt{ViT-L/14@336} configuration, although the contribution of positional embedding resolution cannot be isolated from other differences in the pretrained backbones.

\subsubsection{Physical Appearance vs. Imaging Fidelity}
\label{subsubsec:fidelity}
Evaluating the relationship between anomaly scores and observational parameters establishes an important baseline for interpreting the latent space. Figure~\ref{fig:photometry_results} demonstrates that the median anomaly score remains approximately constant across $0 < z < 6$, and the embedding space in Figure~\ref{fig:photometric_embeddings} shows no correlation with redshift. This lack of dependence may partly reflect the $m_{\rm{F277W}}$ and SNR selection criteria applied to our sample, which preferentially select brighter galaxies at higher redshifts. These selection effects may therefore obscure relationships between anomaly score and redshift-dependent observational limitations, such as cosmological surface brightness dimming and spatial resolution.

Conversely, the representation displays a strong dependence on apparent magnitude and a nonmonotonic relationship with parametric size. Fainter galaxies generally receive higher anomaly scores, while the median score for effective radius peaks near $R_{e} \sim 1.5$ kpc before declining at larger radii (Figure~\ref{fig:photometry_results} and Figure~\ref{fig:parametric_results}). These relationships may partly reflect differences in image fidelity and the amount of resolved structural information available to AdaCLIP. As revealed in Figure~\ref{fig:parametric_embeddings}, larger galaxies frequently display complex substructure (e.g., spiral arms, resolved clumps), whereas smaller galaxies are dominated by single compact components and elevated background noise relative to source flux. The effective-radius gradient in the embedding also broadly overlaps with the apparent-magnitude gradient, with larger galaxies tending to occupy regions associated with brighter sources. This suggests that the embedding may encode related information about galaxy brightness and size, although their individual contributions to the observed spatial organization remain unclear. 

Additionally, the pretrained CLIP representations may encode visual characteristics that are less representative of faint, compact high-redshift galaxies, potentially contributing to elevated anomaly scores even in the absence of intrinsically disturbed morphology. Thus, the model may respond to differences in visual appearance and image fidelity alongside physical structure. To assess whether the observed morphological relationships are primarily driven by galaxy brightness, we examine the anomaly score trends in two magnitude intervals ($20 < m_{\rm{F277W}} < 24$ and $24 < m_{\rm{F277W}} < 28$). The relationships persist in both intervals, suggesting that these trends are not solely driven by the overall $m_{\rm{F277W}}$ dependence. Contaminants within the cutouts such as diffraction spikes, unmasked bright stars, oversized source cutouts, and cutouts with bright edge artifacts, consistently trigger elevated anomaly scores, occupy isolated clusters of the embedding space, or both, confirming that the model reacts strongly to any breakdown in visual uniformity.

\subsubsection{Projected Shape and Structural Asymmetry} 
\label{subsubsec:geometry}
Axis ratio illustrates the value of analyzing the embedding space alongside the scalar anomaly score. The median score rises from highly elongated systems ($q \sim 0.1$) to a peak near $q \sim 0.6$, then plateaus toward round systems ($q \sim 1$) (Figure~\ref{fig:parametric_results}), but the overall variation is small. The low scores of elongated systems may reflect their simple, self-similar geometry, since these are likely edge-on disks, whose projection hides internal structure. The plateau indicates that the model does not rate rounder galaxies as increasingly anomalous beyond $q \sim 0.6$. This may be because the score responds more to visible internal structure than to overall shape: once a disk is not close to edge-on, features such as spiral arms, clumps, and tidal features are visible, and further rounding adds nothing to which the score responds. Despite the similar scores above $q \sim 0.6$, the embedding varies continuously with axis ratio (Figure~\ref{fig:parametric_embeddings}): elongated and round systems occupy visibly distinct regions, with intermediate systems positioned between them. The embedding therefore distinguishes galaxies across the score plateau that the scalar score does not, indicating that it encodes information about galaxy shape beyond what the score captures.

In our sample, median anomaly scores increase steadily with asymmetry (Figure~\ref{fig:CAS_results}), and highly asymmetric systems cluster in a distinct, compact region of the embedding space dominated by mergers and irregular galaxies (Figure~\ref{fig:nonparametric_embeddings}). This strong dependence indicates that AdaCLIP is acutely sensitive to departures from symmetric light distributions.

\begin{figure*}[htb!]
\includegraphics[width=\textwidth]{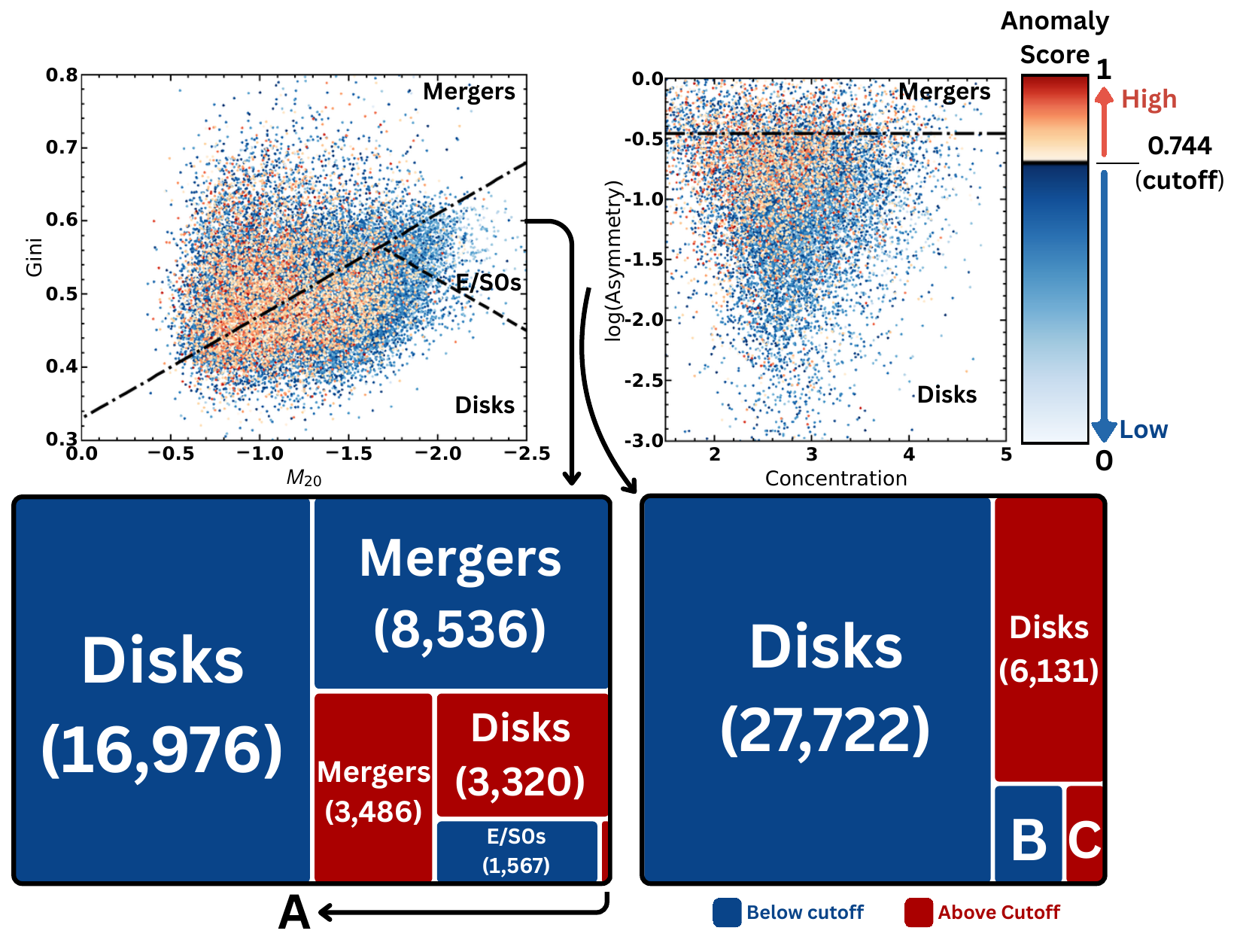}
\caption{Top: rest-frame optical Gini value as a function of $M_{20}$ (left) and rest-frame optical asymmetry as a function of concentration (right). The dashed line separates disks from early-type galaxies (E/S0s), while the dashed-dotted line indicates the locus of mergers ($A = 0.35$; \citealp{Conselice2003}) \citep{Bershady2000, Lotz2008}. Points are colored by anomaly score, with separate gradients applied above and below the anomaly score cutoff (0.744). Bottom: treemap diagrams showing the relative proportions of objects occupying each region in the Gini-$M_{20}$ (left) and asymmetry-concentration (right) planes. The area of each box is proportional to the fraction of the total \texttt{statmorph} sample (33,976 galaxies). Blue and red boxes represent populations below and above the anomaly score cutoff, respectively. Regions too small to display full text are labeled: A = E/S0s above cutoff (91); B = mergers below cutoff (1,357); C = mergers above cutoff (766).}
\label{fig:morphology_plane}
\end{figure*}

\subsubsection{Spatial Organization vs. Flux Statistics}
\label{subsubsec:spatial_vs_flux}
Comparing parametric and nonparametric parameters suggests that AdaCLIP is sensitive to the two-dimensional spatial arrangement of flux, beyond what is captured by one-dimensional radial profiles or non-spatial flux statistics alone. Concentration and $M_{20}$ show systematic relationships with anomaly score: the median anomaly score generally decreases with increasing concentration, whereas it increases with $M_{20}$ before leveling off at the highest values (Figures~\ref{fig:CAS_results} and~\ref{fig:Gini-M20_results}). Both measurements also exhibit spatial organization in the embedding, with localized regions containing bright, centrally concentrated galaxies (Figures~\ref{fig:nonparametric_embeddings} and~\ref{fig:Gini_embeddings}). However, these measurements do not uniquely identify smooth, regular early-type systems. Low-$M_{20}$ galaxies also occupy a separate region containing faint, compact sources, while low-to-intermediate concentration galaxies are distributed broadly throughout the embedding. Thus, while smooth ellipticals form a recognizable population, galaxies with similar nonparametric measurements can occupy different regions of the latent space. This highlights an advantage of AdaCLIP's image-based representation: its ability to distinguish visually different galaxy populations that may otherwise appear similar when characterized by individual nonparametric morphology measurements.

In contrast, Sérsic index and Gini display less straightforward relationships with the model outputs. Although low Sérsic indices trend toward higher anomaly scores, the trend flattens for $n \gtrsim 2$ (Figure~\ref{fig:parametric_results}), and sources across the full range of $n$ remain interspersed throughout the embedding space, apart from a localized population of high-$n$ systems (Figure~\ref{fig:parametric_embeddings}). Similarly, while the Gini coefficient measures pixel flux inequality, it does not account for the spatial locations of those pixels \citep{Lotz2004}. Its relationship with anomaly score is nonmonotonic, with higher median scores at both low and high $G$ and a minimum at intermediate values (Figure~\ref{fig:Gini-M20_results}). Although high-$G$ galaxies show some localization in the embedding, including regions overlapping with low-$M_{20}$ systems, galaxies spanning a range of Gini values remain distributed throughout much of the latent space (Figure~\ref{fig:Gini_embeddings}). Because $M_{20}$ explicitly accounts for the spatial distribution of the brightest regions of a galaxy \citep{Lotz2004}, its more consistent relationship with anomaly score suggests that AdaCLIP is sensitive to the arrangement of bright structures rather than flux inequality alone.

\begin{figure*}[htb!]
\includegraphics[width=\textwidth]{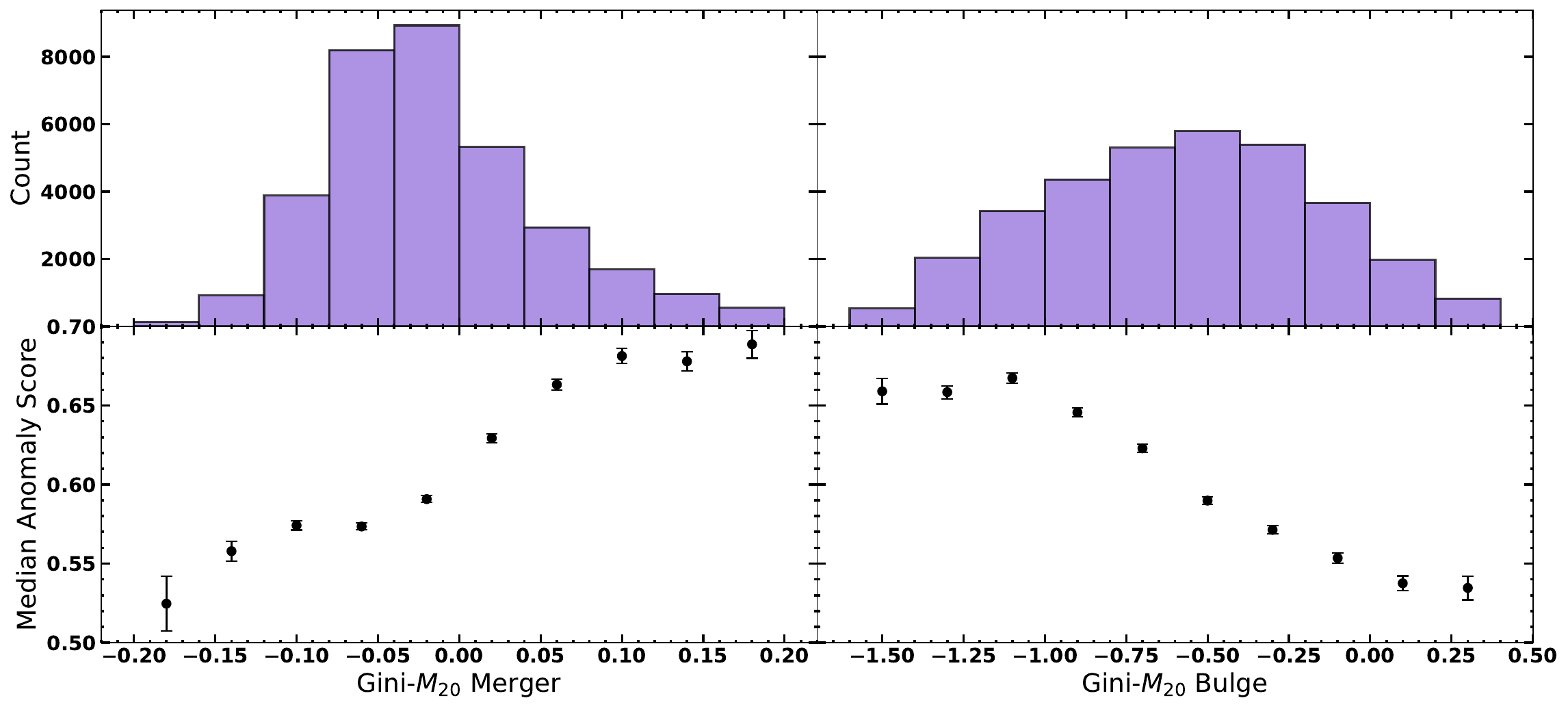}
\caption{Top: Gini–$M_{20}$ merger (left) and bulge (right) parameter distributions of the 33,976 CEERS and NGDEEP galaxies with reliable \texttt{statmorph} measurements. Bottom: median anomaly score within each bin.}
\label{fig:merger_bulge}
\end{figure*}

Overall, zero-shot transfer with AdaCLIP reveals meaningful morphological structure without astronomy-specific anomaly detection training. The model's anomaly scores show modest trends with asymmetry, concentration, $M_{20}$, and size, but weaker or less consistent trends with Sérsic index, axis ratio, smoothness, and Gini coefficient. The feature embeddings reveal additional morphological organization beyond these scalar anomaly score relationships. However, image quality (apparent magnitude and SNR) also influences anomaly scores and the embedding, and should be considered when interpreting these relationships. Together, these results demonstrate the potential of AdaCLIP for identifying both relatively regular systems (e.g., smooth ellipticals and disk galaxies) and structurally disturbed systems (e.g., mergers and other unusual morphologies), while recognizing that observational effects can also contribute to the model's representation.

\begin{figure*}[htb!]
\includegraphics[width=\textwidth]{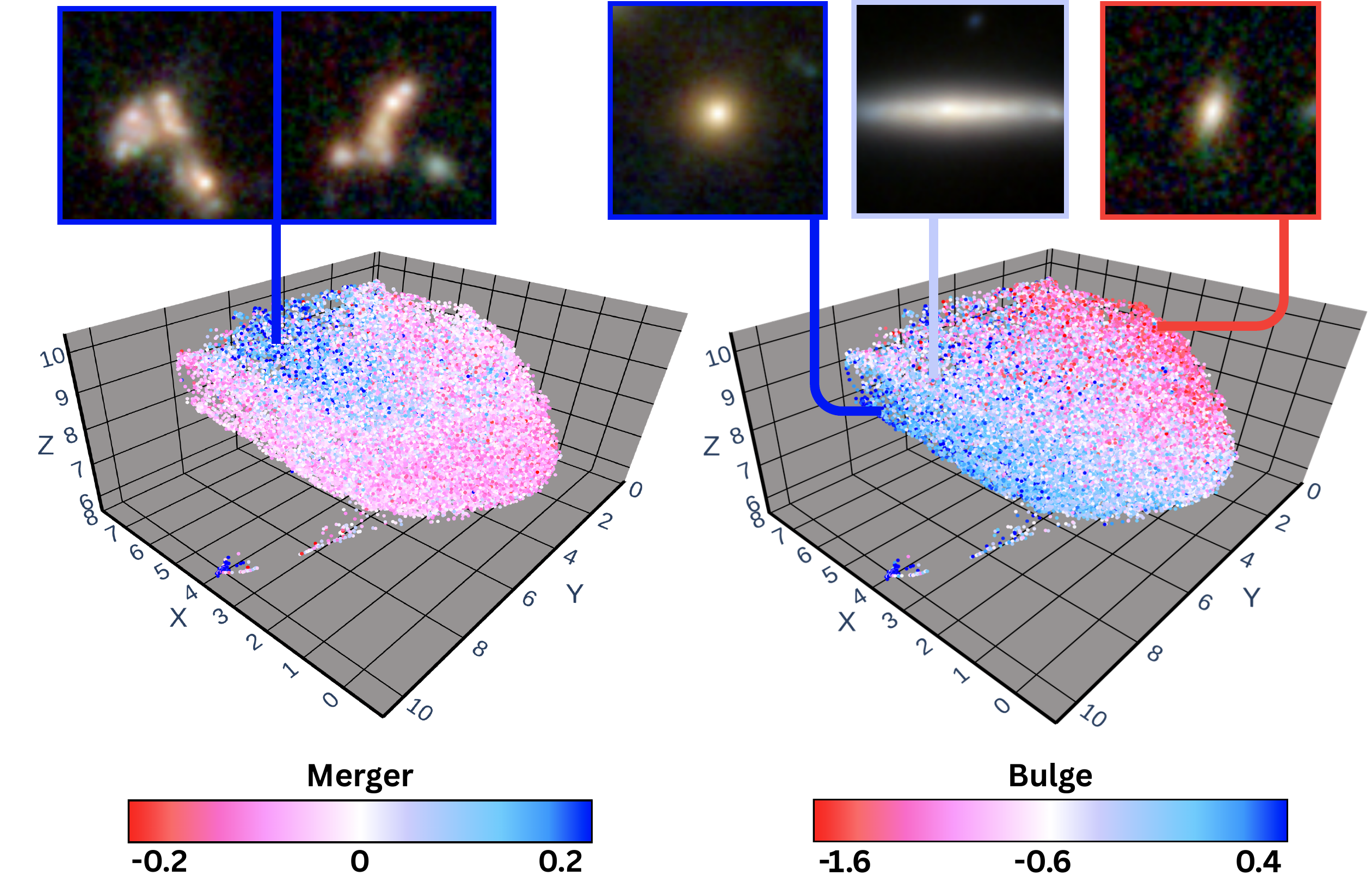}
\caption{Three-Dimensional UMAP projection of galaxies with reliable \texttt{statmorph} measurements, colored by the Gini-$M_{20}$ merger (left) and bulge (right) parameters. Each panel shows the same embedding space with an independent colorbar, which has been rotated to best show groupings or clusters that emerge. Cutouts above each panel are representative galaxies from selected regions of the embedding.}
\label{fig:merger_bulge_embeddings}
\end{figure*}

\subsection{Comparison with Common Diagnostics}
\label{subsec:merging_early_identification}
The top panels of Figure~\ref{fig:morphology_plane} show the combined CEERS+NGDEEP sample in the Gini-$M_{20}$ (left) and asymmetry-concentration (right) planes, with galaxies colored by their anomaly scores. The empirical boundaries in these planes provide conventional reference points for distinguishing morphologically distinct populations \citep{Bershady2000, Lotz2004}. The lines in both planes mark the boundaries between disk, merging, and early-type galaxies based on nearby galaxies at $z \sim 1$ \citep{Lotz2008}.

The distributions in both morphology planes show substantial overlap between galaxies above and below the anomaly threshold. Nevertheless, a modest population-level trend is apparent. As shown in the bottom panels of Figure~\ref{fig:morphology_plane}, $5.5\%$ of galaxies within the E/S0 region lie above the anomaly threshold in the Gini-$M_{20}$ plane, compared with $16.4\%$ within the disk region and $29.0\%$ within the merger region. The asymmetry-concentration plane also shows a similar increase in the fraction of anomalous galaxies toward higher asymmetry. In this plane, $19.3\%$ of galaxies identified as disks have high anomaly scores, while $36.1\%$ of mergers have high anomaly scores. Thus, while the conventional morphology boundaries do not cleanly separate anomalous and non-anomalous galaxies, the fraction of high-anomaly systems increases from regions dominated by early-type galaxies toward those associated with progressively greater morphological disturbance.

The relationship between anomaly score and morphology can be examined more continuously using the Gini-$M_{20}$ merger and bulge statistics \citep{Snyder2015a, Snyder2015b}. These quantities measure the position of a galaxy along directions in the Gini--$M_{20}$ plane associated with increasing merger-like morphology and bulge dominance, respectively. Figure~\ref{fig:merger_bulge} shows the distributions of both statistics together with the median anomaly score in bins of each parameter. The bulge statistic exhibits a pronounced relationship with anomaly score. Galaxies with larger bulge values have lower median anomaly scores. This is consistent with what we see for the concentration and Sérsic indices, as these quantities are strongly correlated \citep{RodriguezGomez2019}.

The merger statistic presents a complementary picture. Consistent with the merger region of the Gini--$M_{20}$ plane containing the greatest fraction of high-anomaly galaxies, the merger statistic exhibits a clear positive relationship with anomaly score. As shown in Figure~\ref{fig:merger_bulge}, the median anomaly score increases from approximately 0.53 at the lowest merger statistic values to 0.69 at the highest values, leveling off toward the upper end of the distribution. Since more positive merger statistic values correspond to increasingly merger-like morphologies, this trend further supports the association between high anomaly scores and structurally disturbed galaxies.

The embedding distributions reveal additional spatial organization associated with both statistics. Galaxies with high merger statistic values occupy a relatively localized region of the embedding (Figure~\ref{fig:merger_bulge_embeddings}), corresponding to the population of visually disturbed and interacting systems previously identified in Figure~\ref{fig:UMAP_CEERS+NGDEEP}. The bulge statistic also exhibits distinct spatial variations, with high values concentrated in a region containing bright, centrally concentrated galaxies. Negative bulge statistic values are associated with more disk-like morphologies, and a less clearly defined grouping of intermediate negative values includes edge-on disks and spiral galaxies. The most negative values, however, are more prevalent in a separate region containing faint, compact sources with substantial background noise. Together, these distributions demonstrate that both statistics exhibit meaningful organization within the AdaCLIP embedding, while also illustrating how the model distinguishes visually different populations across the range of bulge statistic values.

\subsection{Implications and Future Work}
The results of this study demonstrate the potential utility of AdaCLIP as a ZSAD framework for high-redshift galaxy morphology. Because AdaCLIP operates without requiring astronomy-specific fine-tuning or labeled morphological training data, it offers an out-of-the-box solution for exploratory analysis of large, multi-band imaging datasets. As space- and ground-based facilities continue to produce unprecedented volumes of high-resolution imaging, zero-shot architectures provide a scalable mechanism to rapidly identify physically rare phenomena without incurring the overhead of training domain-specific models from scratch. Notably, the configuration selected in this study was adapted using industrial defect and medical colon-polyp imagery, yet its learned representations exhibit meaningful relationships with galaxy morphology, particularly for disturbed and multi-component systems. This demonstrates the potential for computational methods developed using data from seemingly unrelated scientific domains to provide useful insights into astronomical imaging.

Because AdaCLIP requires only a single forward pass to generate both an anomaly score and a high-dimensional feature embedding, it serves as an efficient screening tool, scalable to large surveys. While our current analysis focuses on a quality-selected sample of 38,853 galaxies, this workflow can be readily extended to additional deep JWST imaging fields, such as COSMOS-Web \citep{Scoville2007, Casey2023}, PRIMER \citep{Dunlop2021}, and JADES \citep{Eisenstein2026}. Extending AdaCLIP across these larger datasets will provide a rigorous test of whether the model can consistently separate genuine astronomical sources from faint, noisy, or spurious detections while preserving meaningful structural ordering across expanded samples.  Future work could also extend the analysis to $z>6$, investigating the applicability of zero-shot anomaly detection to earlier galaxy populations whose morphological structures may be increasingly challenging to characterize with the available NIRCam imaging.

An additional consideration for extending this framework is the selection of an appropriate AdaCLIP configuration. Although the \texttt{MVTec-336} model performed best in our evaluation of CEERS and NGDEEP galaxies, its relative performance may depend on the imaging characteristics, galaxy populations, and types of anomalies being investigated. For datasets with comparable NIRCam observations and selection criteria, this configuration provides a reasonable starting point, although similar performance is not guaranteed. For applications involving substantially different datasets, such as observations with different wavelength coverage or imaging characteristics, we recommend repeating the initial model-selection procedure (Pass 1) on a representative subsample before applying AdaCLIP to the full dataset. Further evaluation of the generalizability of different configurations would help establish the extent to which the results of this study can be transferred to other astronomical surveys.

Beyond direct anomaly discovery, the dual AdaCLIP outputs offer a complementary framework for automated sample selection and quality control ahead of computationally intensive structural analyses. Rather than replacing established morphological analysis tools, anomaly scores can serve as a preliminary filter to isolate distinct structural regimes prior to targeted investigation. For instance, because smooth, relaxed systems like ellipticals generally receive lower anomaly scores, studies targeting early massive galaxies could use these scores to preferentially select morphologically regular candidates before performing detailed Sérsic profile fitting.

However, low anomaly scores do not necessarily distinguish undisturbed galaxies from contaminants, as stars can also receive relatively low scores. This consideration is particularly relevant to compact populations such as little red dots (LRDs). A preliminary cross-match with the LRD catalog from \citet{Kocevski2025} identified 47 positional matches within our sample using a $1\farcs0$ matching radius. Although a subset of these LRDs occupies a similar region of the UMAP embedding, many are dispersed throughout the representation. Visual inspection suggests that the clustered LRDs are relatively isolated in their cutouts, whereas the dispersed population frequently contains nearby sources, either physically associated or chance projections, which may influence their positions in the embedding space.

A critical next step following this work is conducting detailed follow-up analyses on the population of high-anomaly objects identified by AdaCLIP. Integrating structural anomaly scores with ancillary data such as spectroscopic observations, multi-wavelength spectral energy distribution fitting, and optical/near-infrared color–color diagnostics will allow us to determine the physical drivers behind high anomaly scores. This multi-wavelength approach will help establish whether visual anomalies correlate with specific evolutionary phases, such as intense starburst events, quiescent systems with unusual structures, obscured active galactic nuclei, or ongoing galactic interactions.

An additional direction for future work is adapting AdaCLIP more specifically to galaxy morphology. While the current implementation uses the general class designation ``galaxy" and generic normal- and abnormal-state phrases, future studies could incorporate morphology-specific classes and state phrases describing undisturbed and disturbed galaxy structures, such as ``symmetric," ``interacting," or ``tidally distorted." \citet{Cao2024} provide training scripts and support for custom datasets, allowing AdaCLIP's framework to be fine-tuned using labeled astronomical images or realistic galaxy simulations. Comparing the anomaly rankings and latent space representations produced by these domain-adapted configurations against those of the original zero-shot model would help determine whether galaxy-specific training improved the identification of visually unusual systems.

Finally, AdaCLIP's anomaly scores and vision embeddings could be integrated into multimodal, large language model-driven scientific analysis frameworks. By connecting a language model to tools for querying catalogs, processing images, running AdaCLIP, and measuring galaxy structure, future systems could accept natural-language requests (e.g., ``find unusual galaxies at $z>6$"), rank candidates by anomaly score, and provide quantitative morphological measurements and descriptions of their structural properties. Such developments could streamline large-scale morphological analyses, particularly with the vast imaging datasets expected from the Vera C. Rubin Observatory's Legacy Survey of Space and Time (LSST) and the Nancy Grace Roman Space Telescope.

\subsection{Caveats}
Several caveats regarding our methodological setup warrant discussion. First, the RGB images used as inputs to AdaCLIP are generated using \texttt{Trilogy}, which applies nonlinear intensity scaling to enhance the visibility of faint structural features. Although identical configuration parameters are applied to all images, the noise and saturation levels used for intensity scaling are determined independently for each cutout. This adaptive scaling may enhance the visual prominence of background noise fluctuations, particularly in images of fainter galaxies, or alter the relative contrast of structural features. Such effects could influence the resulting anomaly scores and potentially contribute to the observed dependence on $m_{\rm{F277W}}$ (Figure~\ref{fig:photometry_results}).

Second, \texttt{GALFIT} and \texttt{statmorph} measure morphological properties from a single NIRCam filter selected to best approximate a rest-frame wavelength of $0.5\mu$m, whereas AdaCLIP operates on RGB composite images constructed from six fixed observed-frame NIRCam filters. Because the rest-frame wavelengths sampled by these filters shift with redshift, the RGB images do not probe a uniform rest-frame wavelength range across the galaxy sample. Furthermore, the single-band images used by \texttt{GALFIT} and \texttt{statmorph} characterize galaxy structure at approximately one rest-frame wavelength, while the AdaCLIP inputs incorporate multi-band color gradients and structural contrasts across a broader wavelength range. Differences in the wavelength-dependent PSFs of the individual filters may also affect the appearance of fine-scale structure in the RGB composites. Consequently, some features contributing to the AdaCLIP anomaly score may not be captured by the single-band morphological measurements, and vice versa, potentially influencing the observed relationships between anomaly scores and conventional morphological diagnostics.

Third, the classification boundaries in the Gini-$M_{20}$ and asymmetry-concentration planes were originally developed and calibrated using $z \sim 1$ baselines \citep{Bershady2000, Lotz2004, Lotz2008}, whereas a substantial portion of our sample extends to $z\gtrsim1$. Cosmological surface brightness dimming \citep{Calvi2014}, spatial resolution limitations, and wavelength-dependent morphological differences can systematically shift galaxies relative to these established criteria. These effects may contribute to the relatively weak separation between high-anomaly and lower-anomaly populations in the merger regions of the respective planes, complicating the interpretation of these diagnostic boundaries for high-redshift galaxies.

Fourth, it should be noted that the F277W SNR and $m_{\rm{F277W}}$ restrictions, together with the redshift limit of $0 < z < 6$, favor a well-detected galaxy population rather than a statistically complete sample. These selections may underrepresent faint, low-surface-brightness, or higher-redshift systems, potentially influencing the distributions of morphological properties and anomaly scores.

Finally, the anomaly score thresholds and UMAP projections used in this study should be interpreted as exploratory tools rather than physically calibrated measures of galaxy morphology. The 80th-percentile anomaly score threshold was selected through visual inspection of galaxy images and is intended to highlight the upper end of each model's anomaly score distribution. Consequently, this threshold is sample-dependent and subject to human judgment, rather than representing an objective boundary between morphologically normal and anomalous galaxies. By construction, the threshold identifies the highest-scoring 20\% of galaxies in a given sample, regardless of the underlying prevalence of physically unusual systems. Therefore, the high-anomaly designation should not be interpreted as a measurement of the true fraction of morphologically anomalous galaxies. Similarly, although the UMAP projections reveal relationships among galaxies in the learned representation space, their projected coordinates and separations do not provide direct physical measurements of morphological similarity.

\section{Summary and Conclusion}
\label{sec:summary_conclusion}
In this work, we obtained anomaly scores and UMAP-projected embeddings for a quality-selected sample of 38,853 galaxies in the CEERS and NGDEEP surveys, spanning $0 < z < 6$ and $20 < m_{\rm{F277W}} < 28$, using the AdaCLIP zero-shot anomaly detection model. We compared these outputs against parametric \texttt{GALFIT} measurements \citep{McGrath2026} and nonparametric \texttt{statmorph} measurements. Our results are summarized as follows.

\begin{enumerate}
    \item  In the model-selection subsample of 1,342 galaxies, the \texttt{MVTec-336} configuration provided the clearest qualitative separation of high-anomaly objects from the broader population. Its embedding distinguished spiral, elliptical, and disturbed systems, including mergers and interacting candidates, while also isolating contaminants such as stars and oversized cutouts.
    \item Applied to the full CEERS+NGDEEP sample, the embedding continues to place high-anomaly objects in a distinct region, dominated by mergers and cutouts containing multiple distinct objects. Stars, noisy detections, oversized cutouts, and sources on the cutout edge form their own groupings. The separation among spiral, elliptical, and edge-on systems is weaker than in the model-selection subsample, consistent with the greater prevalence of compact, small, and faint galaxies in the full sample.
    \item Redshift shows no clear systematic relationship with anomaly score or spatial organization in the embedding. In contrast, apparent magnitude exhibits a smooth gradient across the embedding, with the median anomaly score generally increasing from $\sim 0.36$ for the brightest galaxies to $\sim 0.60$ at fainter magnitudes, before declining slightly in the faintest bin.
    \item Among the parametric measurements, size and axis ratio show clearer relationships with anomaly score than Sérsic index. The median score peaks near $R_e \sim 1.5$ kpc and $q \sim 0.6$, and both quantities vary smoothly across the embedding. Sérsic index shows a slight decline in score with increasing $n$. A weak concentration of high-$n$ galaxies tends toward one edge of the latent space.
    \item Among the nonparametric measurements, $M_{20}$ exhibits a relatively strong change in median anomaly score, increasing from $\sim 0.43$ in the most negative bins to $\sim 0.67$ near $M_{20} \sim -1.0$. Asymmetry shows a smaller increase in median anomaly score, from $\sim 0.54$ to $\sim 0.67$, but more clearly isolates a localized population of highly asymmetric galaxies in the embedding, many of which appear to be mergers or interacting systems. Concentration generally shows decreasing anomaly scores with increasing $C$, with highly concentrated systems occupying a localized region of the embedding. Gini exhibits a nonmonotonic relationship with anomaly score, while smoothness shows only a weak relationship and little apparent organization in the embedding.
\end{enumerate}

Taken together, the trends between anomaly score and morphology are modest and are clearest for spatial and geometric measurements (asymmetry, concentration, $M_{20}$, size, and axis ratio) rather than intensity-profile measurements (Sérsic index, smoothness, and Gini). These results show that a pretrained zero-shot model can identify structurally disturbed high-redshift galaxies and organize them in a way that is broadly consistent with established morphology diagnostics, without domain-specific training.
\begin{acknowledgments}

Support for this work was provided by NASA through grant JWST-ERS-01345.015-A and JWST-GO-02079.021-A awarded by the Space Telescope Science Institute, which is operated by the Association of Universities for Research in Astronomy, Inc., under NASA contract NAS 5-26555. Support for this work was provided by the Rochester Institute of Technology (RIT) Emerson Summer Undergraduate Research Fellowship. This research is based in part on observations made with the NASA/ESA Hubble Space Telescope obtained from the Space Telescope Science Institute, which is operated by the Association of Universities for Research in Astronomy, Inc., under NASA contract NAS 5–26555.

The authors acknowledge Research Computing at the Rochester Institute of Technology for providing computational resources and support that have contributed to the research results reported in this publication. \href{https://doi.org/10.34788/0S3G-QD15}{https://doi.org/10.34788/0S3G-QD15}. 
T.B.S. acknowledges his dog, Bella, who served as an unprompted peer reviewer by repeatedly nudging the keyboard with her protective cone while begging for snacks during late-night writing sessions.

Some of the data presented in this paper are available on the Mikulski Archive for Space Telescopes (MAST) at the Space Telescope Science Institute. The specific observations can be accessed via \dataset[doi:10.17909/z7p0-8481]{https://doi:10.17909/z7p0-8481}.

\facility{JWST}

\software{\texttt{Astropy} \citep{Astropy2013}, \texttt{statmorph} \citep{RodriguezGomez2019}, \texttt{Trilogy} \citep{Coe2012}}

\end{acknowledgments}

\appendix
\section{High and Low Anomaly Score Galaxy Examples}
\label{Appendix:high_low_comparison}
\begin{figure*}[htb!]
\centering
\includegraphics[width=\textwidth]{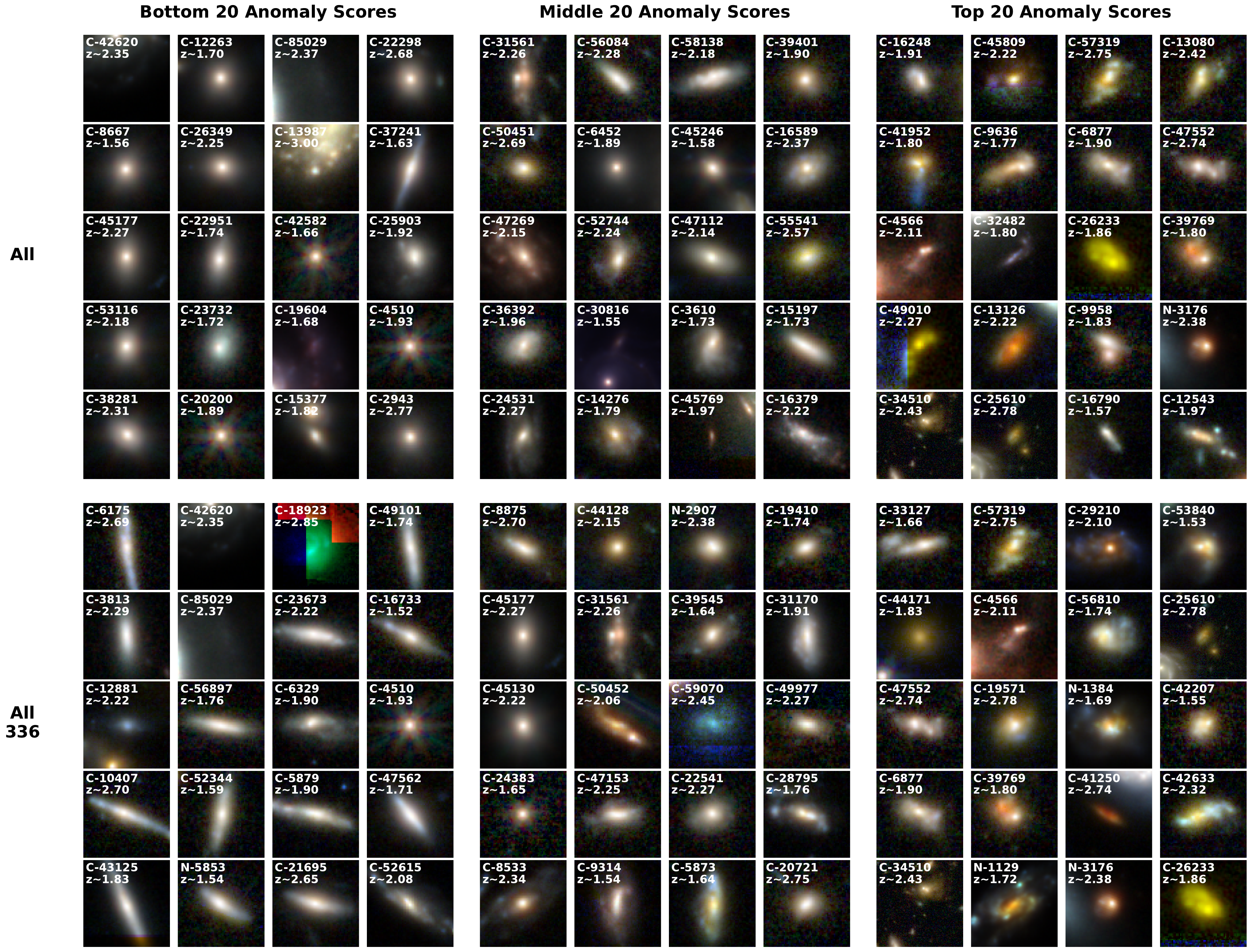}
\caption{The 20 lowest- (left), middle- (center), and highest-scoring (right) galaxy images from the model-selection subsample of 1,342 CEERS and NGDEEP galaxies, ranked by AdaCLIP anomaly score, for the \texttt{All} (top) and \texttt{All-336} (bottom) configurations. The corresponding anomaly score ranges for the six groups, ordered clockwise from the upper-left group, are: 0.032-0.238, 0.566-0.571, 0.852-0.926, 0.881-0.936, 0.689-0.694, and 0.234-0.356.}
\label{fig:low_high_comparison_all}
\end{figure*}

\begin{figure*}[htb!]
\centering
\includegraphics[width=\textwidth]{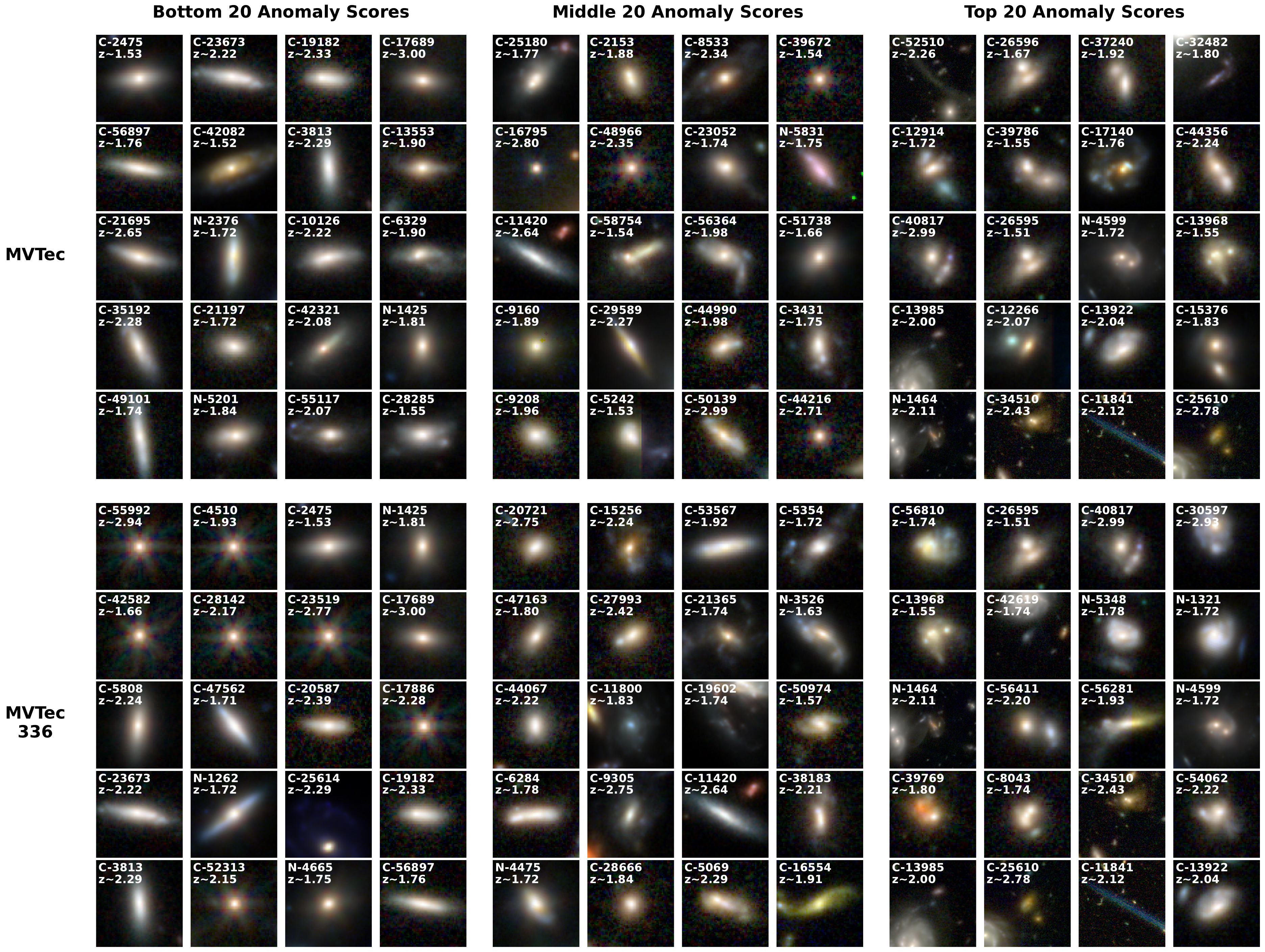}
\caption{The 20 lowest- (left), middle- (center), and highest-scoring (right) galaxy images from the model-selection subsample of 1,342 CEERS and NGDEEP galaxies, ranked by AdaCLIP anomaly score, for the \texttt{MVTec} (top) and \texttt{MVTec-336} (bottom) configurations. The corresponding anomaly score ranges for the six groups, ordered clockwise from the upper-left group, are: 0.057-0.099, 0.334-0.342, 0.803-0.924, 0.872-0.934, 0.473-0.478, and 0.104-0.157.}
\label{fig:low_high_comparison_mvtec}
\end{figure*}

\begin{figure*}[htb!]
\centering
\includegraphics[width=\textwidth]{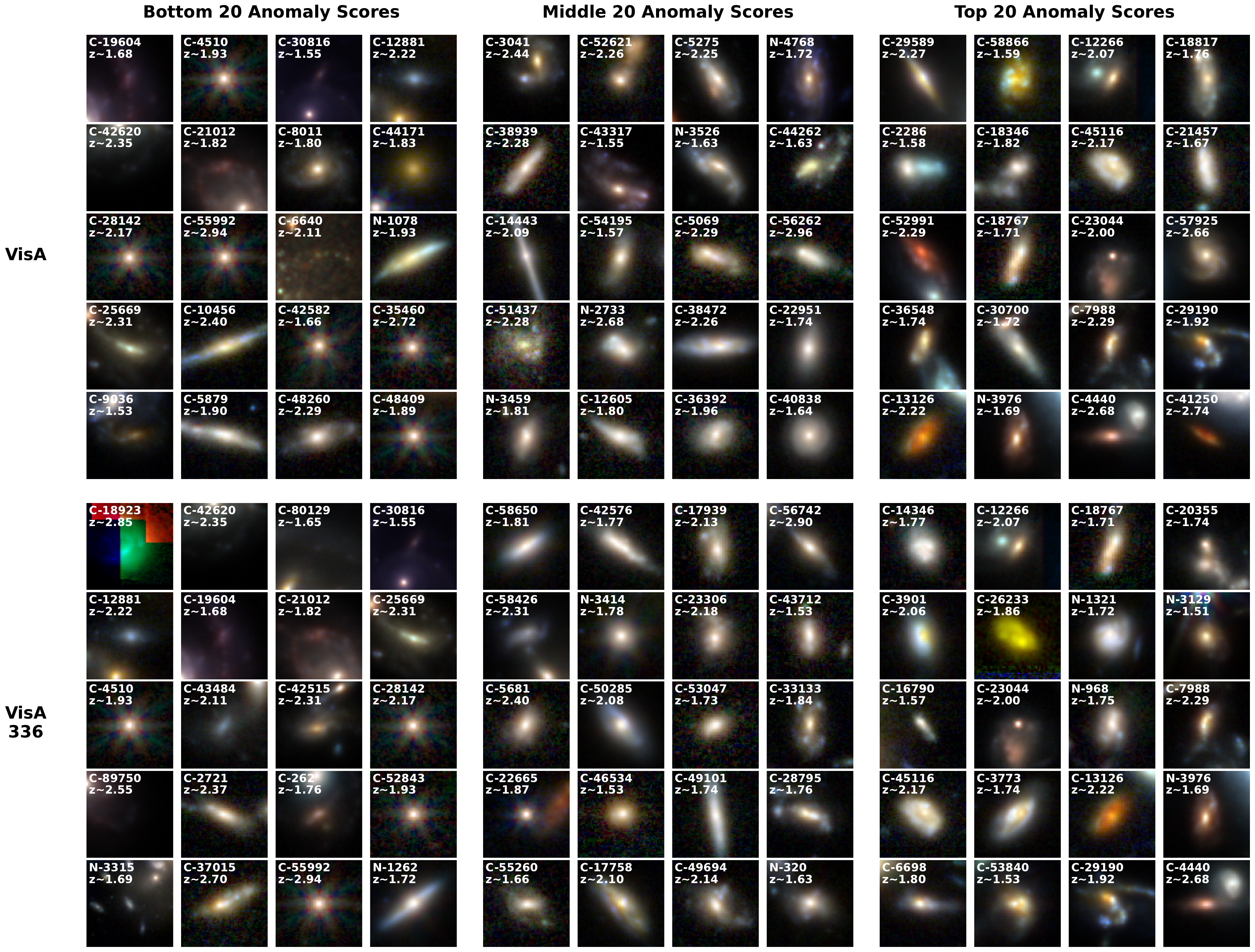}
\caption{The 20 lowest- (left), middle- (center), and highest-scoring (right) galaxy images from the model-selection subsample of 1,342 CEERS and NGDEEP galaxies, ranked by AdaCLIP anomaly score, for the \texttt{VisA} (top) and \texttt{VisA-336} (bottom) configurations. The corresponding anomaly score ranges for the six groups, ordered clockwise from the upper-left group, are: 0.658-0.733, 0.833-0.835, 0.924-0.960, 0.915-0.942, 0.837-0.839, and 0.638-0.748.}
\label{fig:low_high_comparison_visa}
\end{figure*}

As described in Section~\ref{subsec:model_selection}, we show the 20 highest-, middle-, and lowest-scoring galaxies from the subsample of CEERS and NGDEEP galaxies for each of the six model configurations. The examples in Figures~\ref{fig:low_high_comparison_all} - \ref{fig:low_high_comparison_visa} are provided to allow direct visual inspection of the morphologies driving each configuration's anomaly rankings, completing the qualitative comparison presented in Section~\ref{sec:results}. Cutouts in these figures are resized to a common display size for visual purposes and are therefore not to scale with one another. Section~\ref{subsec:image_creation} describes the creation of the cutouts input to the AdaCLIP pipeline.

\section{Embedding Space Comparison}
\label{Appendix:embedding_space}
\begin{figure*}[htb!]
\centering
\includegraphics[width=\textwidth]{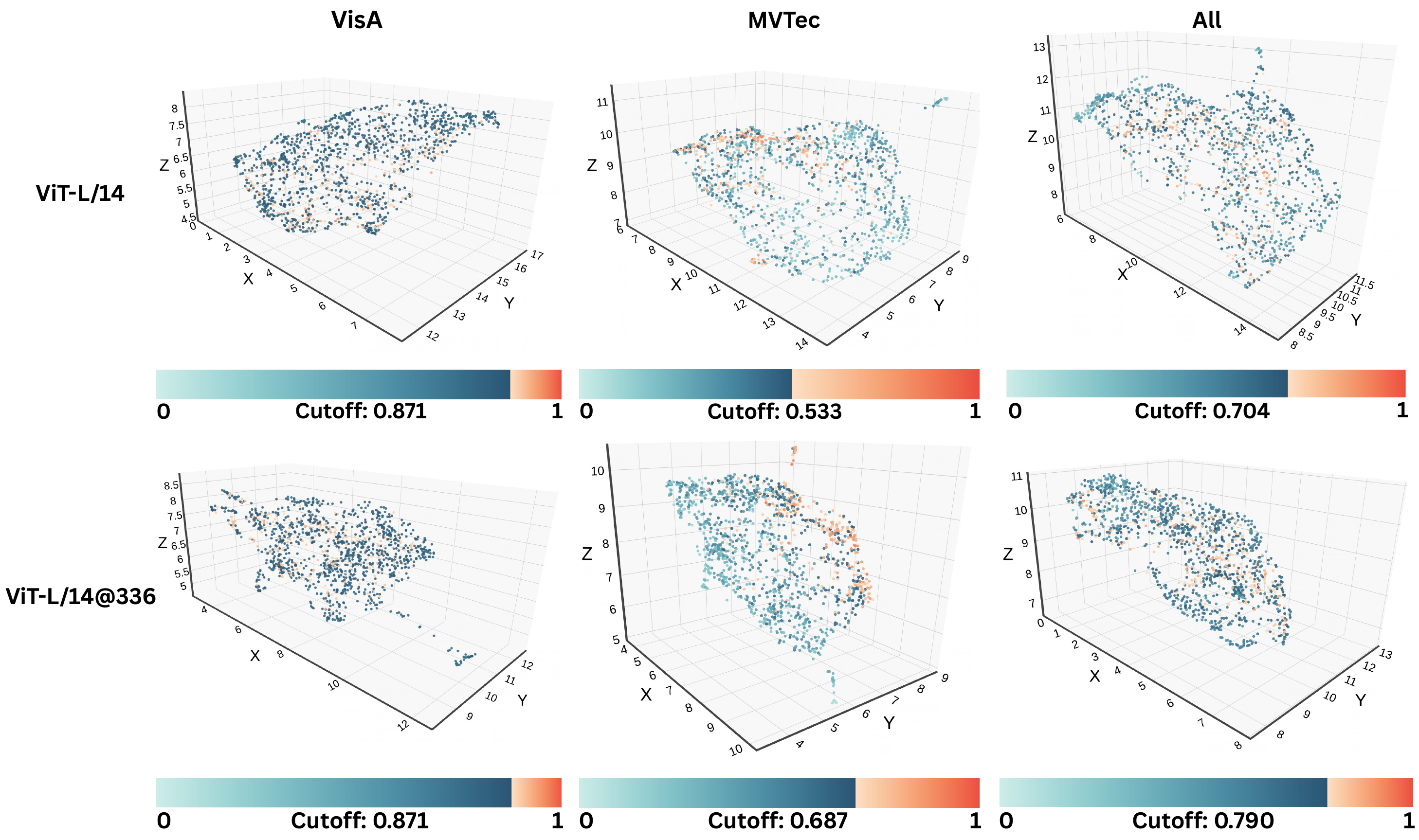}
\caption{Three-dimensional UMAP projections of the AdaCLIP embedding space for all six model configurations, colored by anomaly score. The cutoff value shown in each panel corresponds to the 80th percentile of that configuration's score distribution. Panels are organized by ViT backbone (rows) and AdaCLIP checkpoint configuration (columns).}
\label{fig:six_UMAP}
\end{figure*}

Figure~\ref{fig:six_UMAP} shows the three-dimensional UMAP projections of the AdaCLIP embedding space for all six model configurations, colored by anomaly score. These projections are analyzed for model selection purposes, as described in Section~\ref{subsec:model_selection}. As stated in Section~\ref{subsec:model_selection_results}, we select the \texttt{MVTec-336} model configuration for further analysis. The annotated embedding spaces for the CEERS subset and combined CEERS+NGDEEP sample are shown in Figures~\ref{fig:UMAP_CEERS} and~\ref{fig:UMAP_CEERS+NGDEEP}, respectively.

\clearpage

\bibliography{paper}{}
\bibliographystyle{aasjournalv7}



\end{document}